\documentclass[
 article,
 twocolumn,
 groupedaddress,
 showpacs,
 preprintnumbers,
 amsmath,
 amsthm,
 amssymb,
 aps,
 prstab,
 floatfix,
]{revtex4-1}

\usepackage{graphicx}					
\usepackage{verbatim}					
\usepackage{dcolumn}                    
\usepackage[many]{tcolorbox}
\usepackage[ruled,vlined]{algorithm2e}
\include{pythonlisting}
\usepackage{color}
\usepackage{fontawesome}
\usepackage{adjustbox}

\usepackage{orcidlink}                     

\newcommand{\ds}{\displaystyle}             

\newcommand{\dd}{\mathrm{d}}                
\newcommand{\pd}{\partial}                  

\newcommand{\T}{\mathrm{T}}

\DeclareMathOperator{\sgn}{sgn}             

\newcommand{\const}{\mathrm{const}}

\newcommand{\J }{\boldsymbol{\mathrm{J}}}   
\newcommand{\idt}{\mathbf{I}_2}             

\newcommand{\m}{\mathcal{M}}                
\newcommand{\R}{\mathrm{R}}                 
\newcommand{\K}{\mathcal{K}}                
\newcommand{\h}{\mathcal{H}}                
\newcommand{\pr}{p_r}                       
\newcommand{\pt}{p_\theta}                  

\newcommand{\sn}{\mathrm{sn}}               
\newcommand{\cn}{\mathrm{cn}}               
\newcommand{\dn}{\mathrm{dn}}               
\newcommand{\am}{\mathrm{am}}               
\newcommand{\eK}{\mathrm{K}}                
\newcommand{\eF}{\mathrm{F}}                
\newcommand{\eE}{\mathrm{E}}                

\newcommand{\z}{\zeta}                      

\newcommand{\pmt}{{\color{teal}   \pm}}     
\newcommand{\mpt}{{\color{teal}   \mp}}     
\newcommand{\pmv}{{\color{violet} \pm}}     
\newcommand{\mpv}{{\color{violet} \mp}}     

\begin{document}

\title{Dynamics of McMillan mappings II.    \\
       Axially symmetric map                }
\author{T.~Zolkin\,\orcidlink{0000-0002-2274-396X}}
\email{iguanodyn@gmail.com}
\affiliation{Fermilab, PO Box 500, Batavia, IL 60510}
\author{B.~Cathey\,\orcidlink{0000-0002-6510-9299}}
\email{catheybl@ornl.gov}
\affiliation{Fermilab, PO Box 500, Batavia, IL 60510}
\author{S.~Nagaitsev\,\orcidlink{0000-0001-6088-4854}}
\affiliation{Brookhaven National Laboratory, Upton, NY 11973}
\affiliation{Old Dominion University, Norfolk, VA 23529}
\date{\today}

\begin{abstract}

In this article, we investigate the transverse dynamics of a
single particle in a model integrable accelerator lattice,
based on a McMillan axially-symmetric electron lens.
Although the McMillan e-lens has been considered as a device
potentially capable of mitigating collective space charge
forces, some of its fundamental properties have not been
described yet.
The main goal of our work is to close this gap and understand
the limitations and potentials of this device.
It is worth mentioning that the McMillan axially symmetric map
provides the first-order approximations of dynamics for a
general linear lattice plus an arbitrary thin lens with motion
separable in polar coordinates.
Therefore, advancements in its understanding should give us a
better picture of more generic and not necessarily integrable
round beams.
In the first part of the article, we classify all possible
regimes with stable trajectories and find the canonical
action-angle variables.
This provides an evaluation of the dynamical aperture, Poincar\'e
rotation numbers as functions of amplitudes, and thus
determines the spread in nonlinear tunes.
Also, we provide a parameterization of invariant curves,
allowing for the immediate determination of the map image
forward and backward in time.
The second part investigates the particle dynamics as a function of
system parameters.
We show that there are three fundamentally different configurations of the accelerator optics causing different
regimes of nonlinear oscillations.
Each regime is considered in great detail, including the
limiting cases of large and small amplitudes.
In addition, we analyze the dynamics in Cartesian coordinates
and provide a description of observable variables and
corresponding spectra.
\end{abstract}

\maketitle

\newpage
\section{Introduction}

Understanding the behavior of physical systems is often a
challenging task, requiring the use of numerical simulations and
approximate methods.
However, systems with exact analytical solutions hold a special
place in science.
These rare findings provide important insights into the underlying
principles governing the dynamics of a system.
They serve as benchmarks for testing numerical algorithms and
approximation techniques, allowing us to validate our models and
gain confidence in their predictions.
Moreover, exact solutions offer a deeper understanding of the
fundamental mechanisms at play, shedding light on the interplay
of various factors and revealing hidden symmetries.
While such problems are few and far between, their study paves
the way for a more comprehensive understanding of more general
situations.

Integrable systems, in particular, occupy a central role in the
realm of exact solutions.
They possess a rich mathematical structure that allows for the
explicit determination of their behavior.
The existence of conserved quantities, often referred to as the
{\it integrals} or {\it constants of motion}, provides remarkable
stability and predictability.
This property is particularly valuable in accelerator physics, as
it allows for long-term predictions of the system's dynamics.
Such constants often correspond to physical quantities with
significant importance, e.g. energy, momentum, or angular momentum
and are intimately connected to symmetries.
Exploration of integrability not only provides us with a wealth
of exact solutions, but it also serves as a powerful tool for
understanding the behavior of more complex and non-integrable
cases involving chaotic behavior.

The classical central-force problem emerges as a natural extension
of two fundamental and historically significant integrable systems:
the Kepler problem and the isotropic harmonic oscillator.
The Kepler problem, formulated by Johannes Kepler in the 17-th
century, describes the motion of two bodies under the influence of
gravitational forces and represents one of the earliest examples
of an exactly solvable dynamical system.
The solution by Isaac Newton, published in his monumental work
``Mathematical Principles of Natural Philosophy'' (1687), provided
a physical basis for Kepler's laws and allowed for the derivation
of closed form solutions.
On the other hand, the isotropic oscillator is another cornerstone
since it serves as a fundamental model in various fields of
classical and quantum mechanics.
It lays the foundation for studying harmonic motion and serves as
a basis for various approximation methods.

Remarkably, while a general central-force problem is one of the
most well-studied types of integrable systems, only a select few
potentials yield exact solutions expressible in terms of well-known
functions.
Although numerical methods offer solutions for almost any
central-force problem with arbitrary forces, the existence of
closed-form formulae remains a rarity.
For instance, power-law forces yield analytical solutions in terms
of circular and elliptic functions only when the exponent takes
specific values, such as $1$, $-2$, and $-3$ for circular functions,
and $-7$, $-5$, $-4$, $0$, $3$, $5$, $-3/2$, $-5/2$, $-1/3$, $-5/3$,
and $-7/3$ for elliptic functions, \cite{whittaker1964treatise,
broucke1980notes,mahomed2000application}.
In this article, we present a novel exact solution to another
central-force problem arising from a generalization of the famous
McMillan integrable mapping~\cite{mcmillan1971problem} to higher
dimensions by imposing axial symmetry.
Notably, the resulting system features a biquadratic radial
invariant and a more general Hamiltonian compared to traditional
classical mechanics.

The axially symmetric McMillan map stands out as the only known
exactly integrable nonlinear map in four dimensions that can be
realized in accelerator physics by inserting a specialized nonlinear
electron lens.
The map, initially introduced by R. McLachlan~\cite{McLachlan1993},
underwent further investigation by V. Danilov and E. Perevedentsev~\cite{danilov1997two}
regarding the application of integrable systems to round colliding
beams, with the aim of enhancing the beam-beam limit.
The first practical but approximate concept to realize such a system
was proposed in Ref.~\cite{Danilov1999}.
An electron lens is a device, which uses a low-energy electron beam
to provide nonlinear focusing~\cite{Shiltsev_PhysRevSTAB.2.071001}
for other beams, for example protons or high-energy electrons.
At present, an experimental implementation of such a device is being
developed for the Fermilab IOTA ring~\cite{Antipov_2017}. 

Other known approximate 4D implementations in accelerators of a 2D
McMillan lens are described in Ref.~\cite{PhysRevSTAB.11.114001}.
The only known alternatives are systems based on continuous
(opposite to discrete) dynamics proposed
in~\cite{PhysRevSTAB.13.084002,zolkin2012nonlinear,zolkin2013model}.

More recently, it has been suggested that the axially symmetric
electron lens of McMillan type can be used to mitigate the effects
of space charge (SC) force~\cite{nagaitsev2021mcmillan,
cathey2021calculations,stancari2021beam}.
The linear component of SC force induces a betatron tune shift,
particularly pronounced for particles at zero amplitude, while the
nonlinear component gives rise to an amplitude-dependent betatron
tune spread.
These effects can lead to various undesirable phenomena in
accelerators, such as emittance growth, particle losses, and beam
halo formation.

Although the axially symmetric McMillan map has been known for
some time, a complete analytical solution has remained elusive
until now.
The derived solution not only provides the dynamical variables,
nonlinear tunes, and dynamical aperture but also sheds light on
the possible operational options and different regimes of the lens.
In addition, the system we present in this article holds importance
as it serves as a first-order approximation of an accelerator lattice
incorporating a general axially symmetric nonlinear lens and motion
being separable in polar coordinates.
By considering the expansion of the radial force as a series, it
becomes possible to integrate out the first nonlinear term and obtain
approximated invariant of motion and tunes.
This approximation is particularly valuable in understanding the
behavior of single particle dynamics in complex accelerator structures,
where the inclusion of nonlinear lens elements plays a crucial role.
By studying the properties and dynamics of this simplified integrable
map, we gain insights into the general behavior of particles in 
accelerator lattices with round beams.

\subsection{Article structure}
The structure of our article is as follows:
In Section~\ref{sec:NaturalUnits}, we establish the mathematical
framework and delve into the analysis of the intrinsic parameters
and natural units of the system.
In Section~\ref{sec:RadialMotion}, we focus on the analytical
solution of the radial degree of freedom, while
in Section~\ref{sec:AngularMotion}, we address the angular
motion.
To provide further clarity and detailed derivations,
Appendices~\ref{secAPP:Radial} and \ref{secAPP:Angular} are
included, offering additional insights into specific aspects of
the analytical solutions.
Moving forward, Section~\ref{sec:CriticalCases} examines the
behavior of the system in limiting situations, specifically in
the regimes of large and small amplitudes.
Section~\ref{sec:Regimes} presents a detailed analysis of
different regimes of motion and includes several case studies.
In Section~\ref{sec:RoundBeams} we provide an approximated
invariant for a general round beam lattice with thin nonlinear
electron lens.
Finally, Appendix~\ref{secAPP:Integrals} contains the definitions
of special functions and a comprehensive list of integrals employed
throughout our analysis.
By structuring the article in this manner, we aim to provide a
thorough and coherent exploration of the axially symmetric
McMillan map and its implications in accelerator physics.

\section{\label{sec:NaturalUnits}Natural units}

The most general lattice can be organized by combining a special
linear insert followed by a thin axially symmetric kick, which
represents a short electron McMillan lens.
The linear insert should have equal horizontal and vertical betatron
phase advances and Twiss parameters at the ends, but can otherwise
be arbitrary.
It can be represented by the following matrix equation:
\begin{equation}
\label{math:LinearInsert}
\begin{bmatrix}
    x \\[0.082cm] \dot{x} \\[0.082cm] y \\[0.082cm] \dot{y}
\end{bmatrix}' =
\begin{bmatrix}
  \begin{matrix}
    & & \\[-0.2cm]
    & \mbox{\normalfont\Large\bfseries M} & \\[-0.2cm]
    & &
  \end{matrix}                                              &
  \hspace*{-\arraycolsep}\vline\hspace*{-\arraycolsep}      & \mbox{\normalfont\Large\bfseries 0}                       \\ \hline
  \mbox{\normalfont\Large\bfseries 0}                       &
  \hspace*{-\arraycolsep}\vline\hspace*{-\arraycolsep}      &
  \begin{matrix}
    & & \\[-0.2cm]
    & \mbox{\normalfont\Large\bfseries M} & \\[-0.2cm]
    & &
  \end{matrix}
\end{bmatrix}\cdot
\begin{bmatrix}
    x \\[0.082cm] \dot{x} \\[0.082cm] y \\[0.082cm] \dot{y}
\end{bmatrix},
\end{equation}
where
\[
\mathbf{M} = 
\begin{bmatrix}
    \cos\Phi + \alpha\,\sin\Phi & \beta \,\sin\Phi  \\[0.25cm]
   -\gamma\,\sin\Phi & \cos\Phi - \alpha\,\sin\Phi
\end{bmatrix},
\]
and $\gamma\,\beta-\alpha^2 = 1$ ensures the symplectic condition.
Here, $q=\{x,y\}$ represents the set of transverse Cartesian
coordinates and $\dot{q}=\dd q/\dd s$ represents the corresponding
angular deviations of the particle, where $s$ is the longitudinal
coordinate along the accelerator.
Throughout the rest of the article, the $(')$ symbol will be used
exclusively to denote the application of the map and not a
derivative of any kind.

The nonlinear kick has three parameters, $\mathrm{A}$, $\mathrm{E}$,
and $\Gamma$, but it needs to be ``in tune'' with the rest of the
lattice through its dependence on $\Phi$ and $\beta$.
With the use of polar coordinates $(r,\theta)$, it can be wtitten
in the matrix form as:
\[
\begin{bmatrix}
    x \\[0.082cm] \dot{x} \\[0.082cm] y \\[0.082cm] \dot{y}
\end{bmatrix}' =
\begin{bmatrix}
    x \\[0.082cm] \dot{x} \\[0.082cm] y \\[0.082cm] \dot{y}
\end{bmatrix}
+
\begin{bmatrix}
    0                               \\[0.05cm]
\ds \delta \dot{r}\,\cos\theta      \\[0.05cm]
    0                               \\[0.05cm]
\ds \delta \dot{r}\,\sin\theta
\end{bmatrix},
\]
where
\[
\delta \dot{r}(r) = -\frac{1}{\beta\,\sin\Phi}\,
    \frac{\mathrm{E}\,r}{\mathrm{A}\,r^2+\Gamma} -
    \frac{2\,r}{\beta}\,\cot\Phi.
\]
In experimental setups, this kick can be implemented by inserting
an electron beam into the ring.
Only two parameters remain independent:
\[
    r_m = \sqrt{|\Gamma/\mathrm{A}|},
    \qquad\qquad\text{and}\qquad\qquad
    k_m = \frac{\mathrm{E}}{\Gamma},
\]
representing the characteristic transverse scale of nonlinearity
($r_m$) and the linear focusing strength of the lens ($k_m$),
respectively.
The transverse current density of the beam should follow the
expression:
\[
j_e(r) = \frac{I_e}{\pi\,r^2_m}
\left(
    1 + \mathrm{sgn}\,
    \left[\frac{\Gamma}{\mathrm{A}}\right]\,\frac{r^2}{r_m^2}
\right)^{-2},
\]
that provides an integrated strength of the lens in a thin lens
approximation:
\[
    k_m = \frac{2\,e\,I_e L_m (1-v_e v_p/c^2)}
               {\gamma_p\,m_p\,v_p^2\,v_e\,r_m^2}.
\]
Here, $e$ represents the electron charge, $L_m$ is the length of
the electron beam insertion, $I_m$ denotes the total electron
beam current, $v_{e/p}$ are the velocities of electrons/protons,
$\gamma_p = \sqrt{1-v_p^2/c^2}$ is the Lorentz factor, and $m_p$
stands for the mass of the proton.

The total map exhibits integrability with two functionally
independent invariants, meaning they have a zero Poisson bracket.
These invariants, denoted as $\K_1$ and $\K_2$, can be expressed
as follows:
\[
\begin{array}{l}
\ds \K_1[\dot{x},x,\dot{y},y] = \mathrm{E}\,S(r,\dot{r}) +
        \mathrm{A}\,S^2(r,\dot{r})\,+ \\[0.4cm]
        \ds\qquad\qquad\qquad\qquad +\,\Gamma\,\left[
        r^2 + \frac{S^2(r,\dot{r}) + (\K_2\,\alpha\sin\Phi)^2}{r^2}
        \right] ,                        \\[0.6cm]
\ds \K_2[\dot{x},x,\dot{y},y] = x\,\dot{y} - \dot{x}\,y,
\end{array}
\]
where
\[
    S(r,\dot{r}) = r^2\cos\Phi + r\,\dot{r}\,\alpha\,\sin\Phi
\]
and $r$ and $\dot{r}$ represent the radial coordinate and velocity,
respectively:
\[
    r = \sqrt{x^2 + y^2},
    \qquad\qquad\qquad\qquad
    \dot{r} = \frac{x\,\dot{x}+y\,\dot{y}}{r}.
\]
Before proceeding with the solution, it is necessary to eliminate
any dependent parameters and choose natural units in order to
simplify the further analysis.

(I) As a first step, we can eliminate the parameters related to
the linear lattice.
This can be achieved by performing a transformation to a new set
of coordinates and momenta:
\begin{equation}
\label{math:CanTrans}
\begin{bmatrix}
    q \\[0.25cm] \dot{q}
\end{bmatrix}
\,\rightarrow\,
\begin{bmatrix}
    q \\[0.25cm] p_q
\end{bmatrix} =
\begin{bmatrix}
    1 & 0 \\[0.25cm] \cos\Phi + \alpha\,\sin\Phi & \beta\,\sin\Phi
\end{bmatrix} \cdot
\begin{bmatrix}
    q \\[0.25cm] \dot{q}
\end{bmatrix}.
\end{equation}
This transformation yields the most general ``canonical form'' of
the axially symmetric McMillan map:
\begin{equation}
\label{math:McAxCanCar}
\begin{array}{ll}
\ds q'   &\ds\!\!=   p_q,                             \\[0.35cm]
\ds p_q' &\ds\!\!= - q -
            \frac{\mathrm{E}\,q'}{\mathrm{A}\,(r')^2+\Gamma}.
\end{array}
\end{equation}
In fact, this change of coordinates is equivalent to the original
problem with $\beta = 1$ and $\alpha = 0$ at the location of the
nonlinear lens and $\Phi=\pi/2$.
In polar coordinates
\[
\begin{array}{ll}
\ds x  &\!\!= r\,\cos\theta,                              \\[0.35cm]
\ds p_x&\!\!= \pr\,\cos\theta - \frac{\pt}{r}\,\sin\theta,\\[0.35cm]
\ds y  &\!\!= r\,\sin\theta,                              \\[0.35cm]
\ds p_y&\!\!= \pr\,\sin\theta + \frac{\pt}{r}\,\cos\theta,
\end{array}
\qquad
\begin{array}{ll}
\ds r       &\!\!= \sqrt{x^2 + y^2},                      \\[0.35cm]
\ds \pr     &\!\!= (x\,p_x + y\,p_y)/r,                   \\[0.35cm]
\ds \theta  &\!\!= \arctan(y/x),                          \\[0.35cm]
\ds \pt     &\!\!= x\,p_x + y\,p_y,
\end{array}
\]
the equations of motion for a system can be written in a form
such that the radial and angular degrees of freedom can be
treated independently.
Performing a change of variables provides equations of motion:
\begin{equation}
\label{math:McAxCanPol}
\begin{array}{l}
\ds r'\,= \sqrt{\pr^2 + \frac{\pt^2}{r^2}},             \\[0.4cm]
\ds \pr'= \,-\pr\,\frac{r}{r'} -
\frac{\mathrm{E}\,r'}{\mathrm{A}\,(r')^2+\Gamma},
\end{array}
\quad
\begin{array}{l}
\ds \theta'\,= \theta + \tan\frac{\pt}{r\,\pr},         \\[0.65cm]
\ds \pt'= \pt.
\end{array}
\end{equation}
with corresponding radial
\begin{equation}
\label{math:McAxCanInv}
    \K_r[\pr,r] = \mathrm{A}\,\pr^2\,r^2 + \Gamma\,(\pr^2+r^2)
        + \mathrm{E}\,\pr\,r + \Gamma\,\frac{\K_\theta^2}{r^2}
\end{equation}
and angular invariants
\[
    \K_\theta[\pt,\theta] = \pt.
\]

\newpage
(II) The dynamics of a system is not affected by multiplying an
invariant of motion by a constant value.
After scaling the value of $\K$ and readjusting the parameters of
the map, the equations of motion and the form of the invariants
remain unchanged.
We can achieve this by introducing scaled quantities:
\[
    \widetilde{\K}_r = \frac{\K_r}{\mathrm{A}},
    \qquad\qquad\qquad
    \widetilde{\mathrm{E}} = \frac{\mathrm{E}}{\mathrm{A}},
    \qquad\qquad\qquad
    \widetilde{\Gamma} = \frac{\Gamma}{\mathrm{A}},
\]
With this scaling, the equations of motion (\ref{math:McAxCanCar},
\ref{math:McAxCanPol}) and the invariants (\ref{math:McAxCanInv})
can be expressed in the same form, but with $\mathrm{A} = 1$:
\[
\widetilde{\K}_r[\pr,r] =
    \pr^2\,r^2 +
    \widetilde{\Gamma}\,(\pr^2+r^2) +
    \widetilde{\mathrm{E}}\,\pr\,r +
    \widetilde{\Gamma}\,\frac{\pt^2}{r^2}.
\]

(III) By measuring the Cartesian phase space coordinates in units
of $\sqrt{|\widetilde{\Gamma}|}$, we can introduce scaled
variables:
\[
\begin{array}{l}
\ds (q,p_q)\,\rightarrow\,
    (\overline{q},\overline{p}_q) =
    (q,p_q)/\sqrt{|\widetilde{\Gamma}|},\\[0.25cm]
\ds (r,\pr)\,\rightarrow\,
    (\overline{r},\overline{p}_r) =
    (r,\pr)/\sqrt{|\widetilde{\Gamma}|},\\[0.25cm]
\ds     \pt \,\rightarrow\,
    \overline{p}_\theta =
    \pt/|\widetilde{\Gamma}|.
\end{array}
\]
Furthermore, by performing another rescaling:
\[
    \overline{\K}_r =
        \frac{\widetilde{\K}_r}
        {\widetilde{\Gamma}\,|\widetilde{\Gamma}|},
    \qquad\qquad\qquad
    \overline{\mathrm{E}} =
        \frac{\widetilde{\mathrm{E}}}{\widetilde{\Gamma}},
\]
we effectively eliminate the dependence on the absolute value
of $\widetilde{\Gamma}$.
Introducing a parameter
\[
    a \equiv-\overline{\mathrm{E}} = -\frac{\mathrm{E}}{\Gamma}
\]
and removing all overlines, we arrive at the final form of the
transformation that depends only on one parameter:
\begin{equation}
\begin{array}{ll}
    \ds q'   &\ds\!\!=   p_q,                               \\[0.4cm]
    \ds p_q' &\ds\!\!= - q +
                \frac{a\,q'}{1+\sgn[\Gamma]\,r'^2},
\end{array}
\end{equation}
or in polar coordinates
\begin{equation}
\label{math:McAxCanCar2}
\begin{array}{ll}
    \ds r'   &\ds\!\!=   \sqrt{\pr^2 + \frac{\pt^2}{r^2}},  \\[0.4cm]
    \ds p_r' &\ds\!\!= - \pr\,\frac{r}{r'} +
                \frac{a\,r'}{1+\sgn[\Gamma]\,r'^2},
\end{array}
\quad
\begin{array}{ll}                         
    \ds \theta'&\ds\!\!= \theta + \arctan\frac{\pt}{\pr\,r},\\[0.65cm]
    \ds \pt'   &\ds\!\!= \pt.
\end{array}
\end{equation}

(IV) In the new coordinates, the radial invariant takes the form:
\[
\K_r[\pr,r] =
    \underbrace{\pr^2 - a\,\pr\,r + r^2}_{\propto\mathrm{Courant-Snyder}} +
    \underbrace{\sgn(\Gamma)\,\pr^2\,r^2}_\text{nonlinearity} +
    \underbrace{\pt^2/r^2}_\text{rotation}.
\]
This form of the invariant has a clear physical interpretation for
each term.
The first group of terms, with a combined power of $r$ and $\pr$
equal to 2, represents the linear part of the integral of motion
and is proportional to the Courant-Snyder invariant.
The parameter $a$ corresponds to the trace of the Jacobian
evaluated at the origin, providing the betatron tune:
\[
    \nu_0 = \frac{1}{2\,\pi}\,\arccos\frac{a}{2}.
\]
It's worth noting that the betatron of the lattice is defined
solely based on the parameters of the nonlinear lens and is
independent of $\Phi$.

The second group of terms is responsible for the nonlinearity in
the system.
The sign of $\Gamma$ plays a crucial role and distinguishes two
different types of electron lenses: axially symmetric focusing
and defocusing McMillan octupoles for $\Gamma>0$ and $\Gamma<0$,
respectively.

Finally, the last term can be interpreted as similar to the
kinetic energy of rotation:
\[
    \frac{\pt^2}{2\,m\,r^2}
\]
appearing in the Hamiltonian formulation of the central force
problem.

\section{\label{sec:RadialMotion}Radial motion}

In this section, we focus on the radial motion of the system.
After separating the variables, we obtain a symplectic map of the
plane, which corresponds to an oscillatory system with one degree
of freedom.
In this system, $\pt$ plays the role of a parameter, similar to $a$.
The transformation is invertible and can be expressed as follows:
\[
\begin{array}{lll}
\ds \m_r^{\pmt}:        &
\ds r'\,= \sqrt{\pr^2 + \frac{\pt^2}{r^2}},             &
\ds r>0, \\[0.4cm]
\ds                     &
\ds \pr'= \,-\pr\,\frac{r}{r'} + f(r'),                 &
\ds f(r) = \frac{a\,r}{1 \,\pmt\, r^2},                 \\[0.4cm]
\ds (\m_r^{\pmt})^{-1}: &
\ds r'\,= \sqrt{(f(r) - \pr)^2 + \frac{\pt^2}{r^2}},    &\\[0.4cm]
\ds                     &
\ds \pr'= \quad(f(r) - \pr)\,\frac{r}{r'}.              &
\end{array}
\qquad\qquad\qquad
\begin{array}{l}
\ds r>0,                \\[0.35cm]
\ds f(r) = \frac{a\,r}{1 \,\pmt\, r^2}.
\end{array}
\]
The radial invariant takes the form:
\[
\K_r[\pr,r] =
    \pr^2 - a\,\pr\,r + r^2 \,\pmt\,\, \pr^2 r^2 + \frac{\pt^2}{r^2}.
\]
Here, the plus/minus sign in teal represents one of the two
possible configurations of the electron lens:
the upper/lower sign will always corresponds to $\Gamma = 1$ or
$\Gamma =-1$.
Results specific to only one of the configurations will be labeled
as $[\Gamma_+]$ or $[\Gamma_-]$ respectively.
Since the derivations in both instances are highly similar, we
will present the results concisely, using the colored sign to
distinguish between the two.
In the next subsections, we explore the fundamental properties of
the mappings $\m_r^{\pmt}$, while leaving the details of lengthy
calculations for Appendix~\ref{secAPP:Radial}.

\begin{figure*}[t!]\centering
\includegraphics[width=\linewidth]{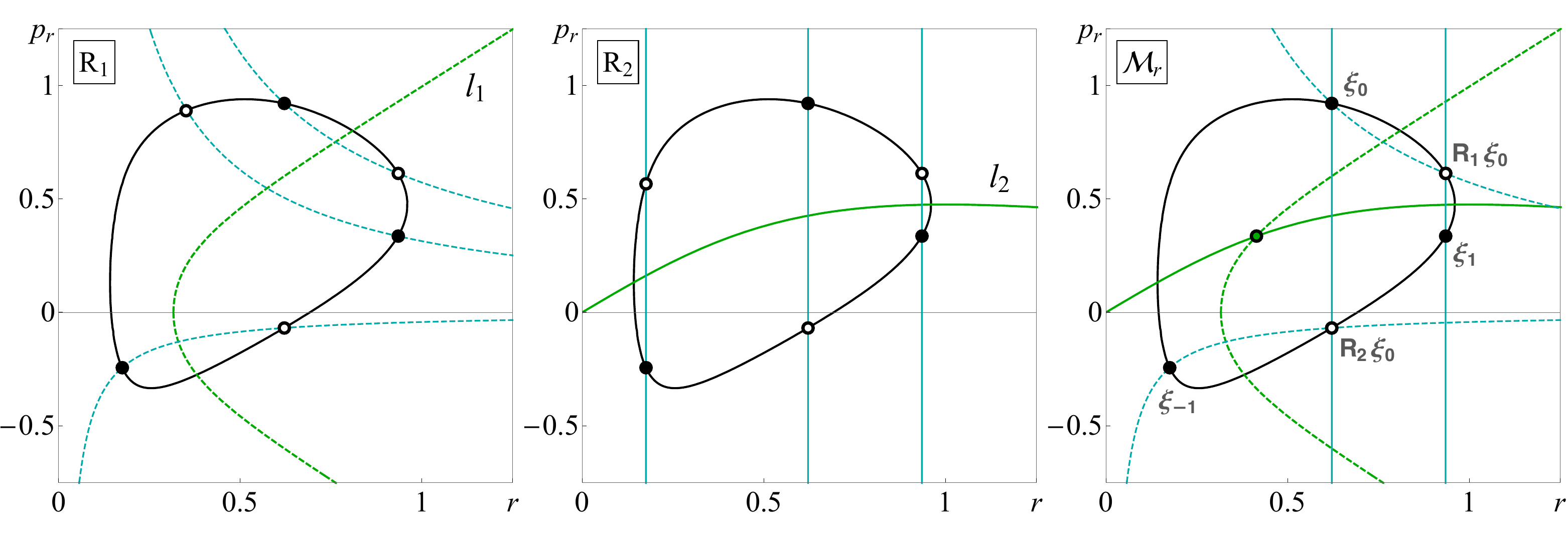}
\vspace{-0.75cm}
\caption{\label{fig:Symmetries}
    Symmetries of the radial part of the axially symmetric McMillan
    map.
    The black closed curve represents a constant level set of the
    radial invariant $\K_r[\pr,r]=1/2$ for the map $\m_r^+$ with
    $a=19/10$ and $\pt = 1/10$.
    The dashed and solid cyan curves correspond to constant level
    sets of $\K_1[\pr,r] = \pr\,r$ and $\K_2[\pr,r] = r$,
    respectively, while the dashed and solid green curves represent
    the first and second symmetry lines, $l_{1,2}$.
    The intersection of $l_1 \cap l_2$ is a fixed point
    of the map.
    The left two plots depict three different initial conditions
    (black dots) and their images (hollow dots) under $\R_{1,2}$.
    The right plot shows the same three points, chosen as consequent
    images $\xi_{-1} \rightarrow \xi_0 \rightarrow \xi_1$ under the
    map $\m_r^+$.
    The hollow dots now represent the iterations of $\xi_0$ under
    $\R_{1,2}$.
    } 
\end{figure*}

\subsection{Symmetry lines}

The direct and inverse mappings of this system can be broken down
into a composition of two nonlinear reflections, denoted by $\R_1$
and $\R_2$, as follows
\[
    \m_r^{\pmt} = \R_2\circ\R_1,
    \qquad\qquad
    (\m_r^{\pmt})^{-1} = \R_1\circ\R_2,
\]
where
\[
\begin{array}{ll}
\ds \R_1: &
\ds r'\,= \sqrt{\pr^2 + \frac{\pt^2}{r^2}},         \\[0.4cm]
\ds              &
\ds \pr'= \quad\pr\,\frac{r}{r'},
\end{array}
\quad\text{and}\quad
\begin{array}{ll}
\ds \R_2: &
\ds r'\,= r,                                        \\[0.5cm]
\ds                     &
\ds \pr'= - \pr + f(r).
\end{array}
\]
This factorization was originally employed by G. D. Birkhoff and
is made possible by the reversibility of the map, please consult
\cite{lewis1961reversible,de1958structure} for the comprehensive
list of references and details.
Both transformations are anti-area preserving involutions, which
means they are their own inverses and their Jacobian determinants
are equal to minus one:
\[
    \R_{1,2} = \R_{1,2}^{-1},
    \qquad
    \R_{1,2}^2 = \idt,
    \qquad
    \det \J_{\R_{1,2}} = -1,
\]
where $\idt$ is the identity matrix.
Each transformation is integrable and has a trivial invariant
\[
    \K_1[\pr,r] = \pr\,r,
    \qquad\qquad\qquad
    \K_2[\pr,r] = r,
\]
satisfying the condition
\[
    \K_i[\R_i(\pr,r)] - \K_i[\pr,r] = 0,
    \qquad\qquad
    i = 1,2.
\]
Moreover, each reflection, and thus their composition, preserves
the radial invariant:
\[
    \K_r[\pr,r] - \K_r[\R_{1,2}(\pr,r)] = 0.
\]
The left two plots in Fig.~\ref{fig:Symmetries} provide an
illustration by showing three different points belonging to a
constant level set of $\K_r$ and their images under $\R_{1,2}$.
Notice how the image of each point again belongs to $\K_r$ and
$\K_{1,2}$, respectively.

Therefore, for both reflections $\R_1$ and $\R_2$, almost all
initial conditions belong to period 2 orbits called 2-{\it cycles},
such that $(r'',\pr'')=(r,\pr)$.
In addition, there are stationary initial conditions satisfying 
$(r',\pr')=(r,\pr)$ and known as {\it fixed points}.
These fixed points form a continuous line of equilibrium
solutions.
The transformation $\R_1$ maps point in phase space with respect
to the line
\[
    l_1:\,\,\pr^2 = r^2 - \frac{\pt^2}{r^2}
\]
while $\R_2$ reflects it vertically with respect to the line
\[
    l_2:\quad\,\,\pr   = \frac{f(r)}{2}.
\]
We refer to these lines as the {\it first} and {\it second
symmetry lines}, respectively.
In particular, it is clear that if $\m_r^{\pmt}$ has any fixed
points, they should belong to the intersection of $l_1$ and $l_2$,
and vice versa (see Fig.~\ref{fig:Symmetries}).

\begin{figure*}[t!]\centering
\includegraphics[width=1\linewidth]{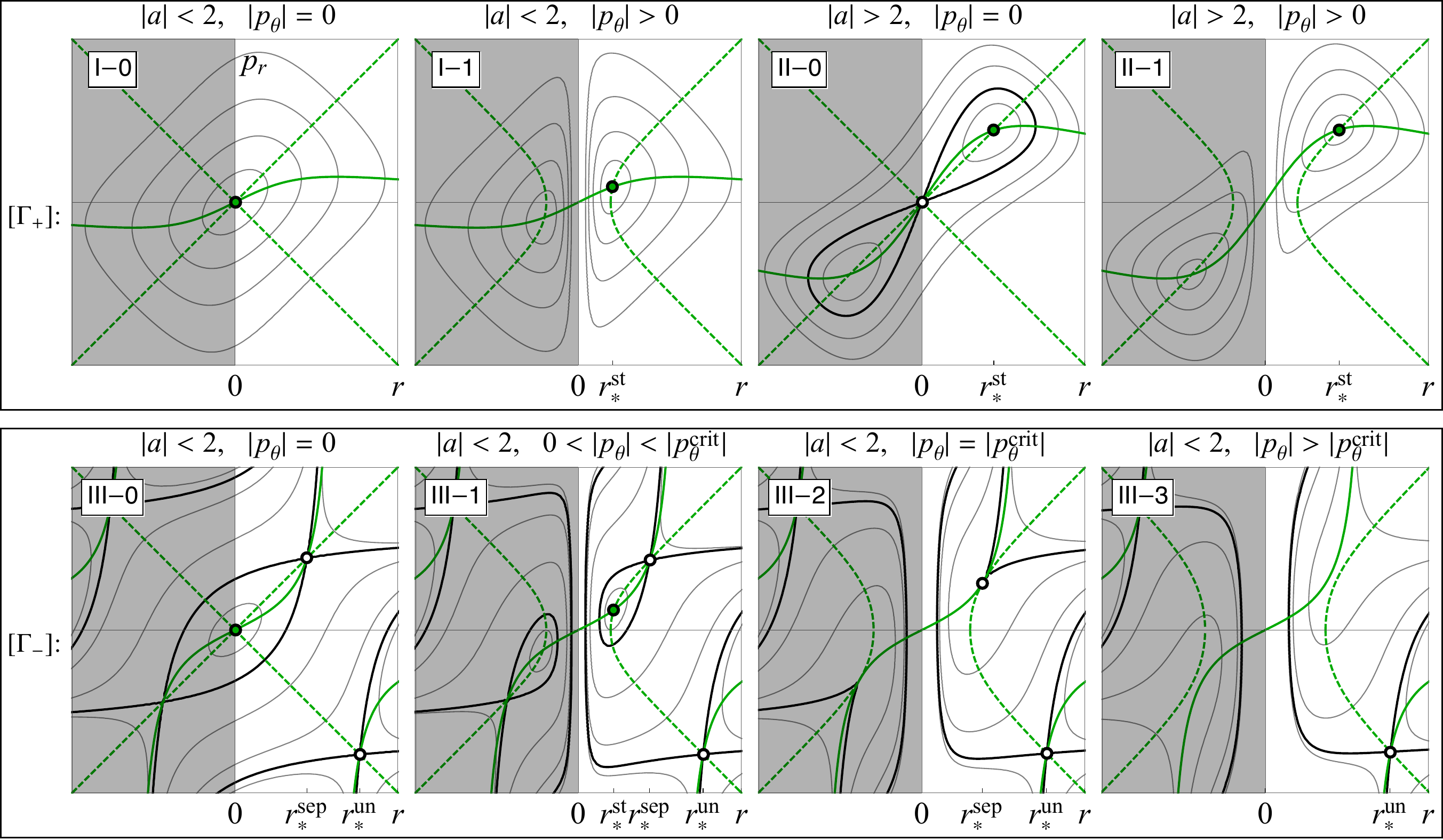}
\caption{\label{fig:KrLevelSets}
    Constant level sets of the radial invariant $\K_r[\pr,r]$ for
    different values of parameters.
    All plots are provided on a square $r,\,\pr\in(-1.6,1.6)$ with
    the nonphysical area of negative radius colored in gray.
    The regular level sets are shown with black curves, while the
    level sets associated with unstable fixed points are shown with
    thick black curves.
    Stable and unstable fixed points of the map are indicated with
    green and white respectively.
    The top and bottom rows are for configurations $[\Gamma_+]$ and $[\Gamma_-]$.
    Roman numerals are used to indicate regimes with possible
    stable trajectories: I and II for $[\Gamma_+]$ with $|a|<2$ and
    $|a|>2$, respectively, and III for $[\Gamma_-]$ with $|a|<2$.
    Sub-cases I-, II-, III-0 represent situations with $\pt = 0$.
    I-, II-, III-1 are for typical situations with stable
    trajectories and $\pt \neq 0$.
    Sub-cases III-2,3 illustrate saddle-node bifurcation and the
    loss of global stability.
    } 
\end{figure*}

The iteration of $\m_r^{\pmt}$ can be interpreted geometrically,
as illustrated in the right plot of Fig.~\ref{fig:Symmetries}.
Starting with any point $\xi_0=(\{r\}_0,\{\pr\}_0)$ on the invariant curve
$\K_r = \const$, we can obtain its image $\xi_1=\m_r^{\pmt}\xi_0$
by first reflecting $\xi_0$ with respect to the first symmetry
line, while ensuring that it remains on the hyperbola
$\pr\,r = \const$ and $\K_r = \const$.
Subsequently, we reflect it vertically ($r=\const$) with respect
to the second symmetry line.
The inverse image of a point can be obtained by reversing the order
of operations, i.e., $\xi_{-1}=\left(\m_r^{\pmt}\right)^{-1}\xi_0$. Additional examples that illustrate the symmetry lines for various
system parameters are shown in Fig.\ref{fig:KrLevelSets}.

\subsection{Fixed points}

Next, let's examine the fixed points of the transformations,
denoted as $(r_*,p_*)$.
They can be found by either using the definition
$(r',\pr')=(r,\pr)$, examining symmetry lines
\[
(r_*,p_*):\,(r_*,p_*)\in l_1 \cap l_2,
\]
or by identifying
the critical points of the invariant~\cite{iatrou2002integrable}:
\[
\left\{\begin{array}{l}
\ds 0=\frac{\pd\K_r}{\pd r  } =
    \,2\, r \,(1\,\pmt\,\pr^2) - a\,\pr,            \\[0.25cm]
\ds 0=\frac{\pd\K_r}{\pd \pr} =
    2\,\pr\,(1 \,\pmt\, r^2) - a\, r - \frac{2\,\pt^2}{r^3}.
\end{array}\right.
\]
This leads to an equation for the roots of the even polynomial of
degree 8:
\[
\mathcal{P}_8(r_*) =
    r_*^8 \,\pmt\,
    2\,r_*^6 +
    \left( 1-\frac{a^2}{4}-\pt^2\right)r_*^4 \,\mpt\,
    2\,\pt^2\,r_*^2 -
    \pt^2 = 0.
\]

The analysis of fixed points, particularly the examination of
their number and stability, provides valuable insights into the
occurrence of bifurcations and the characterization of different
dynamical regimes.
Depending on the configuration $[\Gamma_\pmt]$ and the values of
the parameters, the following scenarios should be considered.

\noindent
$\bullet$ In the case $[\Gamma_+]$, for any values of $a$ and
almost all values of angular momentum (except $\pt = 0$), the
polynomial $\mathcal{P}_8(r)$ has only one positive root.
This root corresponds to the stable fixed point and is denoted
by $r_*^\text{st}$.
Refer to the top row of Fig.~\ref{fig:KrLevelSets}.
We can distinguish two different regimes based on the absolute
value of $a$, with corresponding Roman numerals I and II.

\noindent
$\bullet$ In the case of $[\Gamma_-]$, there are two sub-cases to
distinguish.
For $|a|<2$, we introduce the critical value of the angular
momentum $\pt^\mathrm{crit}$, given by
\[
    \pt^\mathrm{crit} = \left[
        1 - \left( |a|/2 \right)^{2/3}
    \right]^{3/2} < 1.
\]
When $|\pt|<\pt^\mathrm{crit}$, the polynomial 
$\mathcal{P}_8 (r)$ has three positive roots:
\[
    0 < r_*^\text{st} < r_*^\text{sep} < 1 < r_*^\text{un}.
\]
Here, $r_*^\text{sep}$ corresponds to the unstable fixed point
with a separatrix that isolates stable trajectories and
$r_*^\text{un}$ is the second unstable fixed point.
When the absolute value of the angular momentum exceeds the
critical value, the two equilibria $r_*^\text{st}$ and
$r_*^\text{sep}$ collide and annihilate in a saddle-node
bifurcation.
See the last two plots in the bottom row of
Fig.~\ref{fig:KrLevelSets}.
This regime is denoted by the Roman numeral III.
For $|a|>2$, there is only one unstable equilibrium
$r_*^\text{unst} > 1$.
However, since the global dynamics is unstable, we will omit any
further consideration.

\begin{figure*}[t!]\centering
\includegraphics[width=\linewidth]{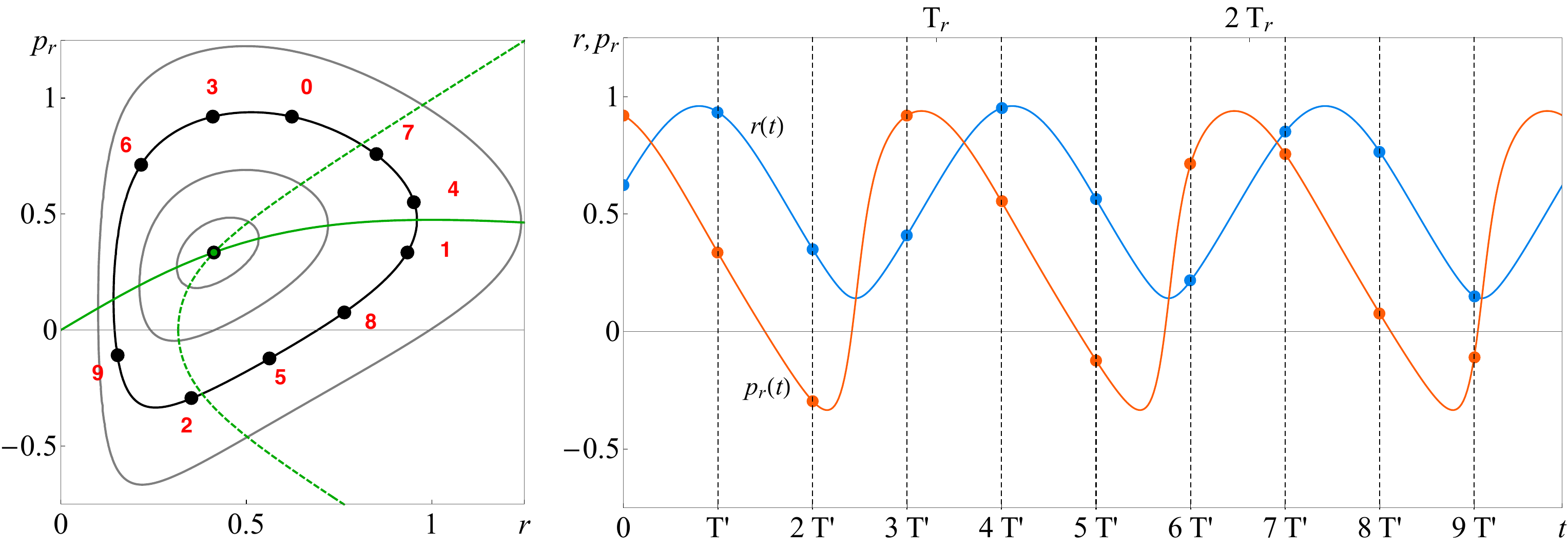}
\caption{\label{fig:RPrParametrization}
    Parametrization of the radial map.
    The left plot shows the constant level sets of the radial
    invariant for the same set of parameters as in
    Fig.~\ref{fig:Symmetries}.
    The level set $\K_r[\pr,r]=1/2$ (shown with a bold curve)
    contains 9 iterations $\xi_i$ of the initial condition $\xi_0$
    (marked with a red numerals $i$).
    The right plot shows continuous solutions $r(t)$ and $\pr(t)$
    of the corresponding Hamiltonian $\K_r[\pr,r;t]$ for initial
    conditions $r,\pr(0) = \{r,\pr\}_0$.
    The discretization of $r(t)$ and $\pr(t)$ at equidistant time
    intervals $t = n\,\mathrm{T}'$ (shown with blue and orange dots) is
    equivalent to the iterations of the map $\m_r^n \xi_0$.
    } 
\end{figure*}

\subsection{Stop points}

We can solve for the radial momentum from the invariant of motion,
which gives us the expression:
\begin{equation}
\label{math:pr}    
\pr =   \frac{f(r)}{2} \,\pmv\,
        \frac{\sqrt{\mathcal{G}_6(r)}}{r\,(1\,\pmt\,r^2)}
\end{equation}
where the violet-colored upper/lower sign corresponds to the solution
on the upper/lower half of the closed invariant curve, respectively.
The polynomial under the square root is given by:
\[
\mathcal{G}_6(r) =
    \mpt\,r^6 -
    \left[ 1 - \left(\frac{a}{2}\right)^2 \mpt\,\,\K_r \right]\,r^4 +
    (\K_r \,\,\pmt\,\, \pt^2)\,r^2-
    \pt^2.
\]
Alternatively, we can introduce a new variable $\z = r^2$ to obtain:
\[
\mathcal{G}_3(\z) =
    \mpt\,\z^3 -
    \left[ 1 - \left(\frac{a}{2}\right)^2 \mpt\,\,\K_r \right]\,\z^2 +
    (\K_r \,\,\pmt\,\, \pt^2)\,\z -
    \pt^2.
\]
For a stable trajectory, the radius is bounded as
$r_- \leq r \leq r_+$ with stopping points belonging to the second
symmetry line.
This means that they have to be solutions of $\mathcal{G}_6(r)=0$.
Depending on the configuration of the nonlinear lens, we have:
\[
\begin{array}{l}
[\Gamma_+]:\quad
\mathcal{G}_3(\z) = (\z_3-\z)(\z-\z_2)(\z-\z_1),\text{ where}   \\[0.25cm]
\quad\z_1 < 0 < \z_2 \leq \z \leq \z_3,
    \quad\qquad
\!r_\mp= \sqrt{\z_{2,3}} = \sqrt{\z_\mp} ,                      \\[0.3cm]
[\Gamma_-]:\quad
\mathcal{G}_3(\z) = (\z_3-\z)(\z_2-\z)(\z-\z_1),\text{ where}   \\[0.25cm]
\quad 0 < \z_1 \leq \z \leq \z_2 < \z_3 < 1,
    \quad
r_\mp= \sqrt{\z_{1,2}} = \sqrt{\z_\mp} .
\end{array}    
\]
Using Vieta's formulas we can also express the map parameter $a$
and invariants as functions of the roots of $\mathcal{G}_3(\z)$:
\begin{equation}
\label{math:ParamZ}
\begin{array}{l}
\ds(a/2)^2 = \prod\limits_{i=1}^3 (1 \,\pmt\,\z_i),         \\[0.35cm]
\ds\pt^2\,= \,\mpt\,\prod\limits_{i=1}^3 \z_i,               \\[0.35cm]
\ds\K_r  =  -\,\prod\limits_{i=1}^3 \z_i \,\mpt\,
    \sum\limits_{\substack{i,j \\ i<j}}^3 \z_i\,\z_j.
\end{array}
\end{equation}

\subsection{Action-angle variables}

The concept of action-angle variables is also can be applied to
symplectic mappings of the
plane~\cite{arnold1968problemes,veselov1991integrable}.
In the action-angle variables, the equations of motion take the
form commonly known as a ``twist map''
\[
\begin{array}{l}
\ds J'\,= J,                 \\[0.3cm]
\ds \psi'= \psi + 2\,\pi\,\nu(J).
\end{array}
\]
Similar to Hamiltonian mechanics, the dynamics of the system can
be decomposed into two parts: the action variables, which remain
constant, and the angle variable, which changes linearly at a
constant rate proportional to the rotation number $\nu$:
\[
    \{\psi\}_n = \{\psi\}_0 + 2\,\pi\,\nu\,n.
\]

For the radial degree of freedom, the action can be expressed in
terms of complete elliptic integrals of the third kind (see
Appendix~\ref{secAPP:Integrals} for definitions of all special
functions).
Specifically, using Eq.~(\ref{math:pr}) we have
\[
\begin{array}{l}
\ds J_r = \frac{1}{2\,\pi}\,\oint \pr\,\dd r =
    \sqrt{\z_3-\z_1}\,\frac{\kappa'^2}{\pi}\,\left\{
    -\,\z_\pmt\,\Pi\left[
        \kappa^2\,\frac{\z_{2\mpt1}}{\z_2},\kappa
    \right]\right.     \\[0.5cm]
\ds \qquad\left.
    \pmt\,(1\,\pmt\,\z_\pmt)\,\Pi\left[
        \kappa^2\,\frac{1\,\pmt\,\z_{2\mpt1}}{1\,\pmt\,\z_2},\kappa
    \right]  \,\mpt\,
    \Pi\left[ \kappa^2,\kappa \right]
    \right\},
\end{array}
\]
where the integral is taken over a closed invariant curve with
$\K_r$ held constant.
The {\it elliptic modulus} $\kappa = \kappa_\pmt$ and
{\it complementary modulus}
$\kappa' \equiv \sqrt{1-\kappa^2} = \kappa_\mpt$ are given in
terms of the roots of $\mathcal{G}_3(\z)$
\begin{equation}
\label{math:kappa}   
    \kappa_+ = \sqrt{\frac{\z_3-\z_2}{\z_3-\z_1}}
    \qquad\mathrm{and}\qquad
    \kappa_-= \sqrt{\frac{\z_2-\z_1}{\z_3-\z_1}}.
\end{equation}


\subsection{Parametrization of the map}

The map can be parametrized using Jacobi elliptic functions.
The expressions for the variables in terms of these functions are
as follows:
\[
\begin{array}{ll}
\ds \{r  \}_n &\ds\!\!\! =
    \sqrt{r_{\pmt}^2 \,\,\mpt\,\, (r_+^2-r_-^2)\,
    \sn^2\left[
        \delta\phi\,n \,\mpt\, \phi_0,\kappa
    \right]},                                               \\[0.45cm]
\ds \{\dot{r}\}_n &\ds\!\!\! = \mpt\,
    \frac{2}{\kappa}\,
    \frac{(r_+^2-r_-^2)^{3/2}}{\{r\}_n}
    \sn\left[ \delta\phi\,n \,\mpt\, \phi_0,\kappa \right]
    \times               \\[0.5cm]
&\ds \quad\times\,
    \cn\left[ \delta\phi\,n \,\mpt\, \phi_0,\kappa \right]\,
    \dn\left[ \delta\phi\,n \,\mpt\, \phi_0,\kappa \right], \\[0.4cm]
\ds \{\pr\}_n &\ds\!\!\! =
    \frac{1}{2}\,\frac{\{\dot{r}\}_n+a\,\{r\}_n}
                      {1\,\pmt\,\{r\}_n^2},
\end{array}
\]
where the phase advance $\delta\phi$ and the initial phase $\phi_0$
can be expressed using complete and incomplete elliptic integrals of
the first kind.
For more information, please refer to Appendix~\ref{secAPP:Radial}.
Specifically, we have
\[
\delta\phi = 2\,\nu_r\,\eK[\kappa],
\qquad
\phi_0 = \pmv\,\eF
    \left[ \arcsin \sqrt{
        \frac{\z_\pmt-\{r\}_0^2}{\z_\pmt-\z_2}
    },\kappa \right],
\]
where $\nu_r$ is the radial rotation number of the map
\[
\nu_r =
    \left\{\begin{array}{ll}
        \mu_r,     & a \geq 0,          \\[0.25cm]
        1 - \mu_r, & a <    0,
    \end{array}\right.
\]
and
\[
    \mu_r = \frac{1}{2\,\eK\left[\kappa\right]}\,
        \eF\left[
            \arcsin \sqrt{\frac{\z_3-\z_1}{1\,\pmt\,\z_\pmt}},
            \kappa
        \right].
\]
The elliptic modulus $\kappa$ for all elliptic functions is
determined by the same equation~(\ref{math:kappa}) as for the
action integral.

Figure~\ref{fig:RPrParametrization} provides an illustration of
the equations mentioned above.
It depicts the parametrization of a specific level set $\K_r$
(left plot) using the continuous functions $r(t)$ and $\pr(t)$
(right plot).
This parametrization is obtained through the solutions of the
corresponding Hamiltonian function $\h[\pr,r;t]$
(see Appendix~\ref{secAPP:Radial}).
The discretization of $r$ and $\pr$ at constant time intervals
$t=n\,\T'$ corresponds to the iteration of the map with initial
conditions given by
\[
\begin{array}{l}
\ds \{r\}_{0\,}\,\,= r(0),     \\[0.35cm]
\ds \{\pr\}_0 = \pr(0).
\end{array}
\]
In this case, the radial rotation number has a clear
interpretation~\cite{zolkin2017rotation,nagaitsev2020betatron}
as the ratio
\[
    \nu_r = \frac{\T'}{\T_r}
\]
where $\T_r$ is the time period of $r(t)$.

\begin{figure*}[t!]\centering
\includegraphics[width=\linewidth]{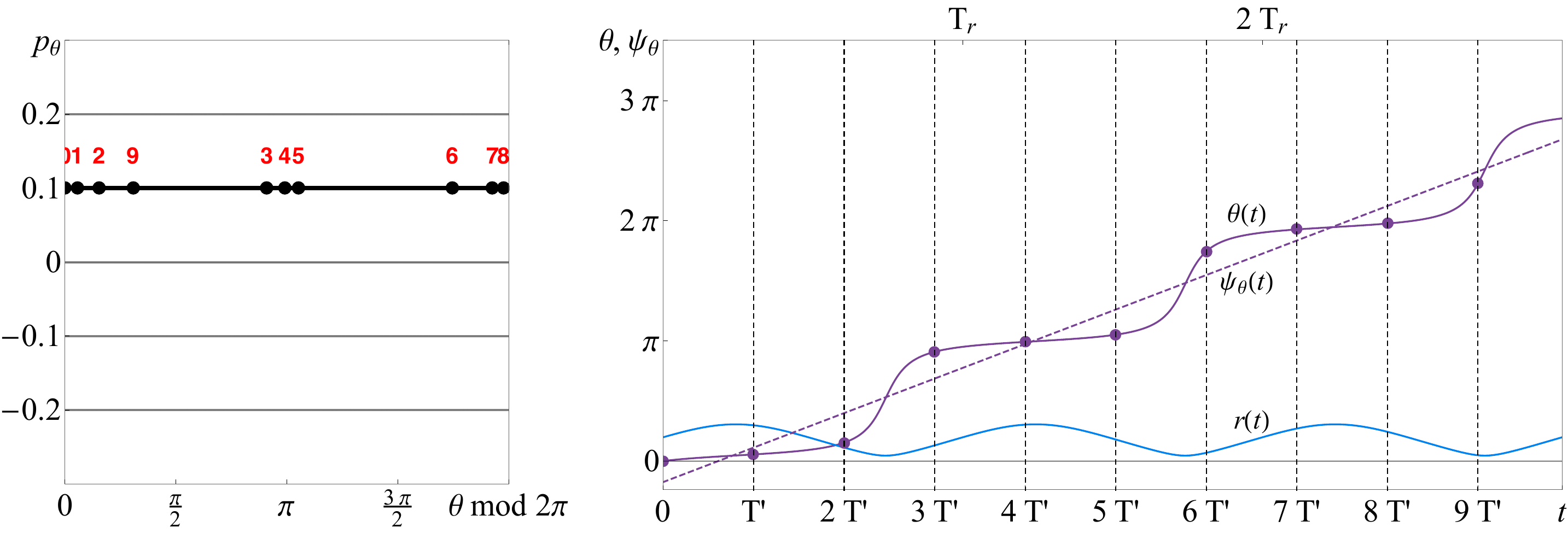}
\caption{\label{fig:ThetaParametrization}
    Parametrization of the angular map $\m_\theta$.
    The left plot shows the constant level sets of the angular
    invariant.
    The level set $\pt=1/10$ (shown with a bold curve) contains 9
    iterations for the same set of initial conditions as in
    Fig.~\ref{fig:RPrParametrization} and $\{\theta\}_0 = 0$.
    The right plot shows the continuous solution for the angular
    coordinate $\theta(t)$ and the canonical angle variable
    $\psi_\theta(t)$.
    The discretization of $\theta(t)$ at the same rate $\T'$ is
    equivalent to the iterations of the map $\m_\theta$.
    } 
\end{figure*}

\section{\label{sec:AngularMotion}Angular motion}

In this section, we present analytical results for the angular part
of the map, while the actual derivations are provided in Appendix
\ref{secAPP:Angular}.
The mapping equations are given by:
\[
\begin{array}{ll}
\ds (\m_\theta)^{\pm 1}:   &
\ds \pt'= \pt,          \\[0.35cm]
\ds                     &
\ds \theta'= \theta \pm \arctan\frac{\pt}{r\,\pr}.
\end{array}
\]
It is evident that $\pt$ is invariant, and hence, the study of
dynamics is essentially reduced to a non-autonomous circle map
that has explicit dependence on the iteration number through the
radial variables.
The angular variable can be parametrized with the help of Jacobi's
amplitude, $\am$, as follows:
\[
\begin{array}{l}
\ds \{\theta\}_n =  \{\theta\}_0 +
                \{\delta\Theta\}_n - \{\delta\Theta\}_0 +
                (\Delta_\theta'-\Delta_\Theta')\,n,         \\[0.35cm]
\ds \{\delta\Theta\}_n =
    \frac{\pt}{\z_\pmt\,\sqrt{\z_3-\z_1}}\,
    \Pi\left[
        \am\left[
            \delta\phi\,n \,\mpt\, \phi_0,
            \kappa
        \right],
        1-\frac{\z_\mpt}{\z_\pmt},
        \kappa
    \right],
\end{array}
\]
where
\[
\begin{array}{l}
\ds \Delta_\Theta' =
    \left\{\begin{array}{ll}
        \Delta_\mu,                 & a \geq 0,     \\[0.35cm]
        \Delta_\Theta - \Delta_\mu, & a <    0,
    \end{array}\right.                      \\[0.65cm]
\ds \Delta_\theta'\,= 
    \arctan\left(
        \frac{2\,\pt}{a}\,\frac{1\,\pmt\,\z_\pmt}{\z_\pmt}
    \right) +
    \pi\,\mathrm{sgn}[\pt]\,\mathrm{H}[-a], \\[0.45cm]
\ds \Delta_\Theta = 
\frac{2\,\pt}{\z_\pmt\,\sqrt{\z_3-\z_1}}\,
    \Pi\left[
        1-\frac{\z_\mpt}{\z_\pmt},
        \kappa
    \right],                                \\[0.35cm]
\ds \Delta_\mu =
\frac{\pt}{\z_\pmt\,\sqrt{\z_3 - \z_1}}\,
    \Pi\left[
        \arcsin \sqrt{\frac{\z_3-\z_1}{1 \,\pmt\, \z_\pmt}},
        1-\frac{\z_\mpt}{\z_\pmt},
        \kappa
    \right].
\end{array}
\]

The solution is obtained by discretizing the arithmetic
quasiperiodic function $\{\theta\}_n = \theta(n\,\T')$ (see Fig.
\ref{fig:ThetaParametrization}), which can be written as a sum of
periodic and linear functions
\[
\theta(t) = \theta_\mathrm{per}(t) +
    \frac{
        \nu_r\Delta_\Theta + \Delta_\theta' - \Delta_\Theta'
    }{\T'}\,t,
\]
such that
\[
\forall\,t:\,\,
    \theta_\mathrm{per}(t+\T_r) = \theta_\mathrm{per}(t).
\]
The linear advancement of the angular coordinate is equal to that
of the canonical angle variable, defining the angular rotation
number, $\nu_\theta$.
In terms of action-angle variables, the angular map can be written
as:
\[
\begin{array}{l}
\ds J_\theta'\, = J_\theta,                 \\[0.5cm]
\ds \psi_\theta'= \psi_\theta + 2\,\pi\,\nu_\theta,
\end{array}
\qquad\qquad
\begin{array}{l}
\ds J_\theta   = |\pt|,                     \\[0.3cm]
\ds \nu_\theta = \nu_r\,\frac{\Delta_\Theta}{2\,\pi} +
        \frac{\Delta_\theta'-\Delta_\Theta'}{2\,\pi},
\end{array}
\]
where the action variable associated with the angular motion is
equal to the absolute value of the angular momentum.

\section{\label{sec:CriticalCases}Critical cases}

To understand all possible regimes and associated modes of
oscillations, we begin the detailed investigation of dynamics by
analyzing the critical cases that correspond to large and small
values of the action variables.

\subsection{Large radial amplitudes, $r \rightarrow \infty$}

As we have seen, in the case $[\Gamma_+]$, the dynamical aperture
is unbounded.
Therefore, we can consider the limit of large radial amplitudes.
In this situation, the force function tends to zero:
\[
    \lim_{r\to\infty} f(r) = 0.
\]
Thus, the effect from nonlinearity vanishes, and the resulting 2D
harmonic oscillator possesses two different types of degeneracy.
First of all, the motion is decoupled in $x$ and $y$, so the system
is separable not only in polar coordinates
\[
\begin{array}{l}
\ds    r'\,     = \sqrt{\K_r^{(\infty)} + \pt^2}/r, \\[0.35cm]
\ds    \pr'     =-\pr\,r/r',
\end{array}
\]
but also in Cartesian coordinates $(q=x,y)$:
\[
\begin{array}{l}
\ds    q'\, = \,\,p_q,      \\[0.4cm]
\ds    p_q' = - q,
\end{array}
\]
with two functionally independent sets of invariants:
\[
    \K_r^{(\infty)} = \pr^2\,r^2, \quad \pt = \const,
\qquad\mathrm{and}\qquad
    \K_q = p_q^2 + q^2.
\]

\begin{figure*}[t!]\centering
\includegraphics[width=\linewidth]{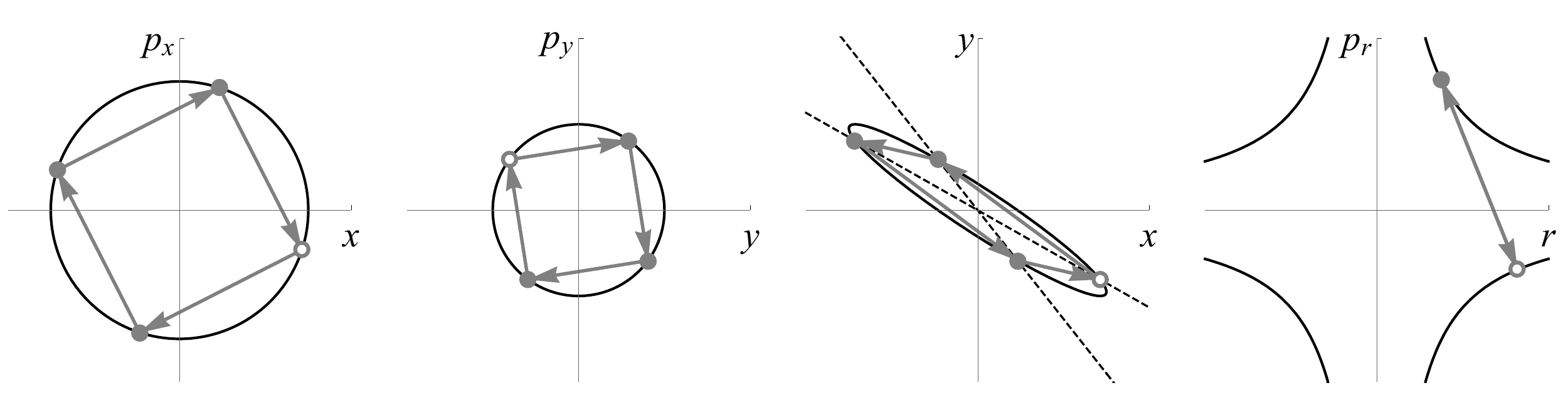}
\vspace{-0.5cm}
\caption{\label{fig:2dLinearOscillator}
    Dynamics for large radial amplitudes.
    The left two plots show the Cartesian invariant $\K_q=p_q^2+q^2$
    in $(q,p_q)$-planes, the third plot illustrates the projection
    onto the $(x,y)$-plane (Lissajous oval), and the last plot
    shows the radial phase space with the limiting invariant
    $\K_r^{(\infty)} = \pr^2\,r^2$.
    Iterations of the map are marked with gray points (initial point
    with white core) connected by arrows.
    } \vspace{-0.5cm}
\end{figure*}

\noindent
The solution to the Cartesian degrees of freedom is given by
\[
\begin{array}{lll}
\ds\{q  \}_n                                        & \ds \!\!=
\pmv \,q_+ \sin
    (2\,\pi\,\nu_0^{(\infty)}\,n \,\pmv\, \phi_0),  \\[0.35cm]
\ds\{p_q\}_n                                        & \ds \!\!=
\{q  \}_{n+1},
\end{array}
\]
where
\[
\begin{array}{lll}
\ds q_+ = \sqrt{\K_q},                              \\[0.35cm]
\ds\phi_0 = \arcsin(\{q\}_0/q_+),
\end{array}
\]
and $\nu_0^{(\infty)} = 1/4$.
While the first degeneracy is related to the fact that $\nu_x=\nu_y$,
an additional super-degeneracy appears due to the fact that the
Cartesian frequencies $\nu_{x,y} = \nu_0^{(\infty)}$ are rational.
In fact, any function such that
\[
    \K [q,p_q] = \K [-q,p_q] = \K [q,-p_q]
\]
is a Cartesian invariant of motion, for example $p_q^2\,q^2$ in
addition to $p_q^2 + q^2$.

During the iterations, the particle visits four distinct points in
the $(q,p)$-planes
\[
\begin{array}{l}
\ds ( \{q  \}_0, \{p_q\}_0 )      \,\rightarrow\,
    ( \{p_q\}_0,-\{q  \}_0 )      \,\rightarrow\,
    (-\{q  \}_0,-\{p_q\}_0 )      \\[0.25cm]
\ds\qquad\rightarrow\,
    (-\{p_q\}_0, \{q  \}_0 )      \,\rightarrow\,
    ( \{q  \}_0, \{p_q\}_0 )      \,\rightarrow\,
    \ldots,
\end{array}
\]
and two distinct points in the $(r,\pr)$-plane
\[
  ( \{r \}_0, \{\pr\}_0 )                   \,\rightarrow\,
  ( \frac{\sqrt{\K_r^{(\infty)} + \pt^2}}{\{r\}_0},
   -\{\pr\}_0  \frac{\{r\}_1}{\{r\}_0} )    \,\rightarrow\,
  \ldots.
\]
After 4 iterations, the particle advances in angular coordinate by
$2\,\pi$:
\[
\begin{array}{l}
\ds\{\theta\}_0                       \,\rightarrow\,
  \{\theta\}_0 + \delta\theta        \,\rightarrow\,
  \{\theta\}_0 + \pi                 \\[0.25cm]
\ds\qquad\quad\rightarrow\,
  \{\theta\}_0 + \delta\theta + \pi  \,\rightarrow\,
  \{\theta\}_0 + 2\,\pi,
\end{array}
\]
where
\[
\delta\theta
    = \arctan\frac{\pt}{\{r \}_0\,\{\pr\}_0}
    = \arctan\frac{\pt}{\sqrt{\K_r^{(\infty)}}}.
\]
Notably, we have the following relations between the fundamental
frequencies:
\[
    \nu_{x,y} = \nu_\theta = \nu_r/2 = \nu_0^{(\infty)}
\]
which have a clear geometrical interpretation.
Points in the $(x,y)$-plane belong to a Lissajous ellipse:
\[
\left(\frac{y}{\{y\}_0}\right)^2 -
2\,\frac{y}{\{y\}_0}\,\frac{x}{\{x\}_0}\,\cos\delta\phi +
\left(\frac{x}{\{x\}_0}\right)^2 =
\sin^2\delta\phi_0,
\]
where
\[
\delta\phi_0 = \phi_0^{(y)} - \phi_0^{(x)},
\]
and for which the radius oscillates twice per complete revolution,
as illustrated in Fig.~\ref{fig:2dLinearOscillator}.

\begin{figure*}[t!]\centering
\includegraphics[width=\linewidth]{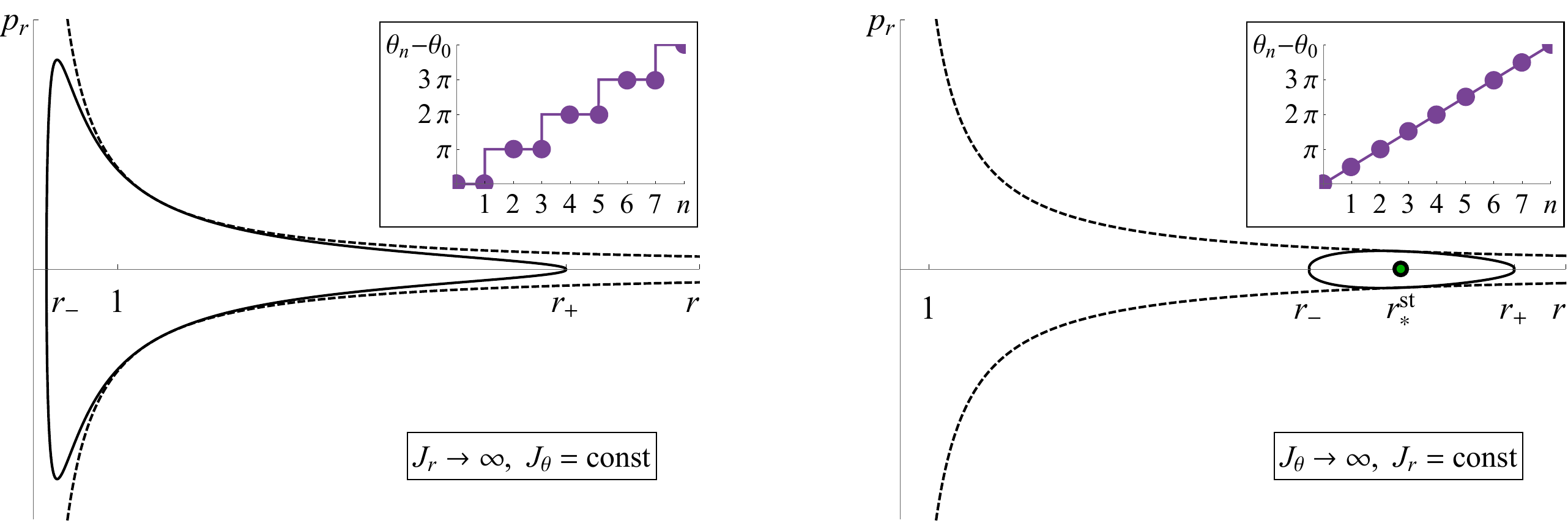}
\vspace{-0.25cm}
\caption{\label{fig:KrLargeR}
    Deformation of closed invariant curves towards the limiting
    invariant $\K_r = \pr^2\,r^2$ (dashed curves) as
    $J_r\rightarrow\infty$ with constant angular momentum and
    $J_\theta\rightarrow\infty$ with fixed $J_r$.
    Both plots are for the case $[\Gamma_+]$.
    Additional plots show limiting behavior of angular part of
    the map $\{\theta\}_n - \{\theta\}_0$ (purple dots) along
    with corresponding parametrization $\theta(t)-\theta_0$.
    } 
\end{figure*}

It is worth mentioning that there are two different ways in which
large radial amplitudes can be achieved: either $J_r \rightarrow
\infty$ while $\pt$ is constant, or $J_\theta \rightarrow \infty$
for a fixed value of $J_r$.
The corresponding approximations of the mapping equations are:
\vspace{-0.25cm}
\[
\begin{array}{ll}
J_r \rightarrow \infty\,:\quad
&\ds    r'\,     \approx \frac{\sqrt{\K_r^{(\infty)}}}{r} = |\pr|, \\[0.35cm]
&\ds    \pr'     \approx-r\,\sgn\pr,
\end{array}
\]
and
\[
\begin{array}{ll}
J_\theta \rightarrow \infty\,:\quad
&\ds    r'\,     \approx \frac{|\pt|}{r} =
    \frac{(r_*^\mathrm{st})^2}{r}, \\[0.35cm]
&\ds    \pr'     \approx-\pr\,\frac{r}{r'}.
\end{array}
\]
When $J_r\rightarrow\infty$, the radial map is equivalent to a
"fold" of the Cartesian linear map, where the absolute value function
and sign guarantee that $r$ stays positive.
The area under the invariant curve must increase with $J_r$, causing
it to occupy the entire space under the limiting invariant
$\K_r^{(\infty)} = \pr^2\,r^2$.
The minimum possible radius $r_-$ tends to zero, and the maximum
radius $r_+$ goes to $\infty$ as $J_r$ increases:
\[
    0 < r_- \ll 1 \ll r_+.
\]
The shape of the invariant curve resembles a T-handle of a cane or
wrench, as shown in the left plot of Fig.~\ref{fig:KrLargeR}.
In the second case, the fixed point exhibits the following limits:
\[
    \lim_{\pt\rightarrow\infty} r_*^\mathrm{st} = \sqrt{|\pt|},
    \quad
    \lim_{\pt\rightarrow\infty} \pr^\mathrm{st} = 0,
    \quad
    \lim_{\pt\rightarrow\infty} \K_*^\mathrm{st} = 2\,|\pt|.
\]
As $J_r$ remains fixed, the closed invariant curve is ``pushed''
towards the right as $\pt\rightarrow\infty$ (see the right plot in
Fig.~\ref{fig:KrLargeR}), resulting in the condition
$
1 \ll r_\pm.
$
Consequently, the radial motion becomes more uniform and resembles
circular orbits.

\newpage
Another way to comprehend this difference is by examining the
angular DOF.
In the limit of the linear oscillator, we have:
\[
\{\theta\}_n = \sgn\pt\,
\arctan\left[
    \frac{r_+}{r_-}\,
    \tan\left(
        2\,\pi\,\nu_0^{(\infty)}\,n \,\pmv\, \phi_0
        \right)
\right]
\]
or in terms of action-angle variables
\[
\begin{array}{l}
\ds \{\theta\}_n = \sgn\pt\,\times          \\[0.25cm]
\ds \times\arctan\left[
    \sqrt{\frac{2\,J_r+J_\theta+2\,\sqrt{J_r\,(J_r+J_\theta)}}
               {2\,J_r+J_\theta-2\,\sqrt{J_r\,(J_r+J_\theta)}}}\,
    \tan\frac{\{\psi_r\}_n}{2}
\right].
\end{array}
\]
Applying the limits, we obtain:
\[
\lim_{J_r\rightarrow\infty}\delta\theta = 0
\qquad\mathrm{and}\qquad
\lim_{J_\theta\rightarrow\infty}\delta\theta = \pi/2.
\]
Hence, we observe that in the former case, $\theta$ only jumps by
$\pi$ every second iteration, while in the latter case, it increases
by $\pi/2$ each time, resulting in the uniform rotation discussed
earlier.
These solutions can be seen as the discretization of the floor and
linear functions:
\[
    \mathrm{floor}\,[2\,\pi\,\nu_0^{(\infty)}\,n,\,\pi]
    \qquad\mathrm{and}\qquad
    2\,\pi\,\nu_0^{(\infty)}\,n
\]
respectively, see Fig.~\ref{fig:KrLargeR} for illustration.

\subsection{Zero angular momentum, $J_\theta = 0$}

Next, we consider the critical case of zero angular momentum,
$\pt = 0$.
The motion in the $(x,y)$-plane is constrained to a line defined
by $\{\theta\}_0 = \mathrm{const}$ and essentially becomes
one-dimensional.
Denoting the Cartesian coordinate along this line as $q$, we observe
that the mapping equations for the radial DOF correspond to a folded
(to keep $r$ positive) one-dimensional {\it octupole} or {\it
canonical} McMillan map.
We can compare the equations as follows:
\[
\begin{array}{l}
\ds    r'\,    = |\pr|, \\[0.35cm]
\ds    \pr'    =-r\,\sgn\pr + \frac{a\,r'}{1\,\pmt\,r'^2},
\end{array}
\quad\mathrm{vs.}\quad
\begin{array}{l}
\ds    q'\,    = p, \\[0.35cm]
\ds    p'    =-q + \frac{a\,q'}{1\,\pmt\,q'^2}.
\end{array}
\]
Furthermore, the corresponding invariants are given by:
\[
\begin{array}{l}
\ds\K_r[\pr,r]\,= \pmt\,\pr^2\,r^2 + \pr^2 - a\,\pr\,r + r^2,    \\[0.25cm]
\ds\K_\mathrm{oct}[p,q] = \pmt\,p^2\,q^2 + q^2 - a\,p\,q + q^2.
\end{array}
\]
This dynamical system has been extensively studied in
\cite{mcmillan1971problem,iatrou2002integrable,zolkin2022mcmillan,
zolkin2025MCdynamics}.
Here, we will mention only the qualitative features that are
important for the discussion, and we encourage the reader to refer
to the aforementioned references for more details.

\begin{figure}[p!]\centering
\includegraphics[width=0.85\linewidth]{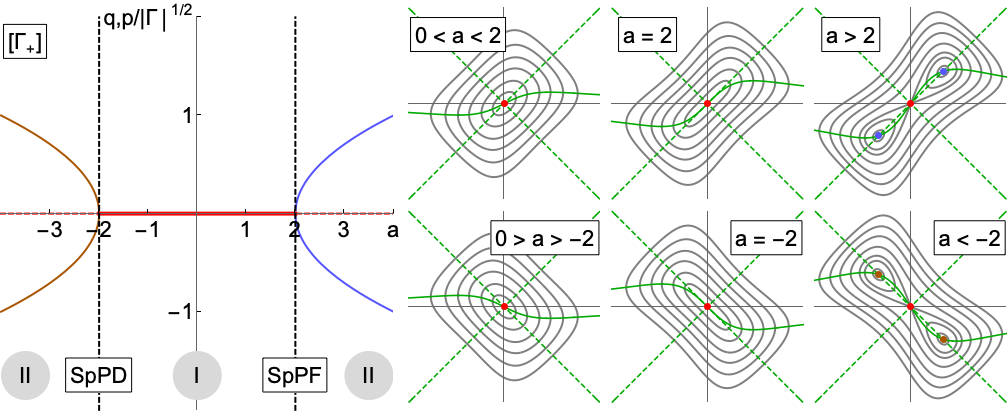}
\vspace{-0.25cm}
\caption{\label{fig:1DMcMGammaP}
    Fixed points and 2-cycles of the 1D McMillan octupole map
    $\m_\mathrm{oct}^\mathrm{1D}$ for the case $[\Gamma_+]$.
    The figure on the left shows the control plot illustrating
    the coordinate/momenta of critical points, scaled by
    $\sqrt{|\Gamma|}$.
    Stable fixed points (unstable fixed points) are represented
    by solid (dashed) lines.
    The fixed point at the origin is shown in red, symmetric fixed
    points in blue, and the 2-cycle is depicted in brown.
    Plots on the right show invariant level sets $\K_\mathrm{oct}[p,q]$
    for different values of the parameter $a$.
    } 
\end{figure}

$\bullet$ In the case $[\Gamma_+]$, when $a=2$ (or $a=-2$), the
system undergoes a supercritical pitchfork (supercritical period
doubling) bifurcation.
The fixed point at the origin $(q,p)_*^{(1)} = (0,0)$ becomes
unstable, and an additional pair of stable symmetric fixed points
(or 2-cycle) emerges:
\[
(q,p)_*^{(2,3)} = \pm\sqrt{(-2 + a)/2}\,(1, 1).
\]
Fig.~\ref{fig:1DMcMGammaP} illustrates the control plot for the
fixed points and the 2-cycle of the 1D octupole McMillan map,
highlighting both bifurcations in the parameter space.
The plots on the right show corresponding constant level sets of
the invariant $\K_\mathrm{oct}$ for different values of $a$.
For $|a|>2$, there exist two distinct modes of oscillations,
which are separated by a figure-eight shaped separatrix.
Trajectories outside the separatrix encircle the unstable fixed
point located at the origin, while trajectories inside the
separatrix revolve solely around the symmetric fixed points or
2-cycle.

\onecolumngrid

\begin{figure}[p!]\centering
\includegraphics[width=0.85\linewidth]{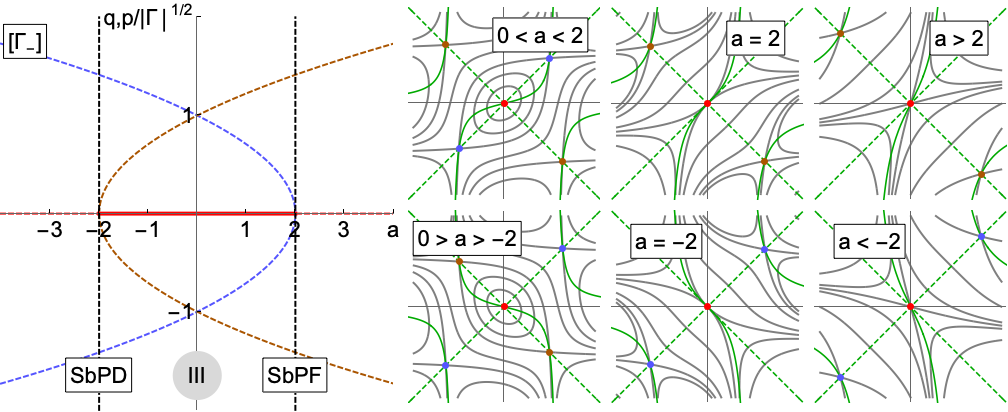}
\vspace{-0.25cm}
\caption{\label{fig:1DMcMGammaM}
    Same as in Figure~\ref{fig:1DMcMGammaP}, but for the case
    $[\Gamma_-]$.
    } 
\end{figure}

\begin{figure}[p!]\centering
\includegraphics[width=0.95\linewidth]{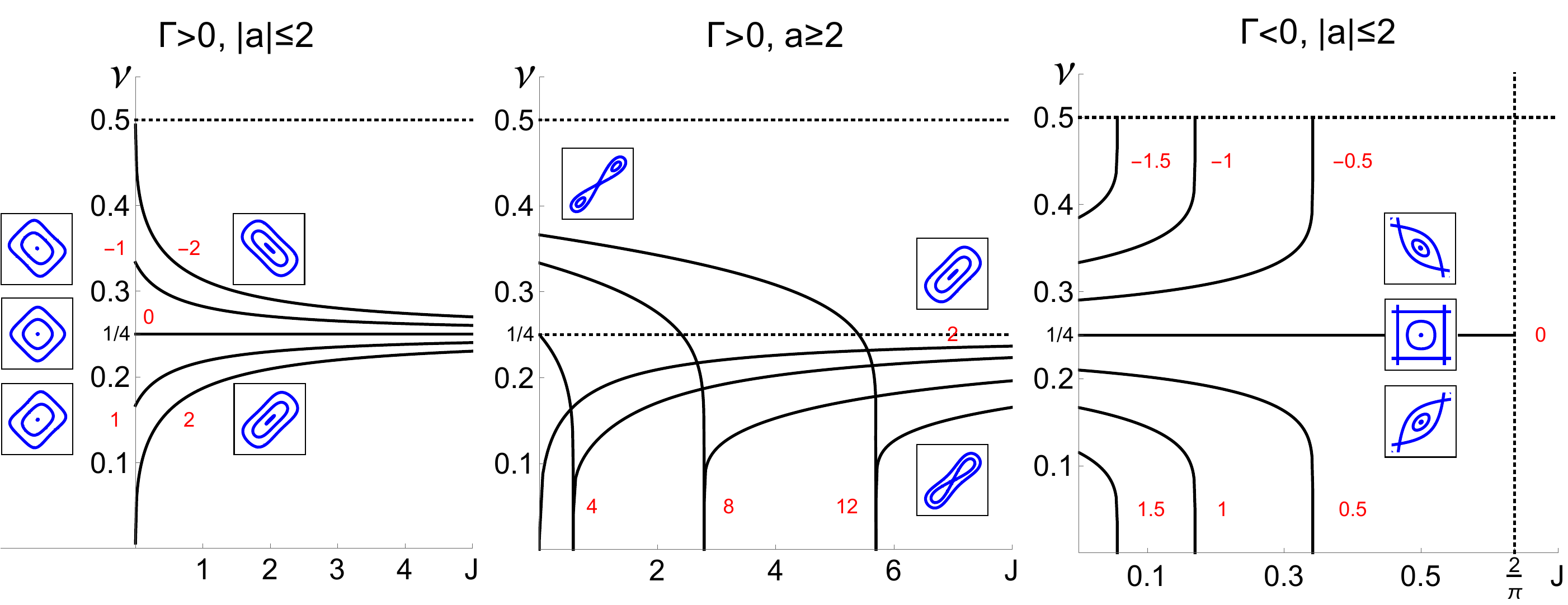}
\vspace{-0.25cm}
\caption{\label{fig:JNuOct}
    Rotation number as a function of the action variable for the
    1D McMillan octupole map.
    Different curves correspond to different values of $a$,
    indicated by the red labels.
    Adapted from~\cite{zolkin2022mcmillan,zolkin2025MCdynamics}.
    } 
\end{figure}

\begin{figure}[t!]\centering
\includegraphics[width=0.95\linewidth]{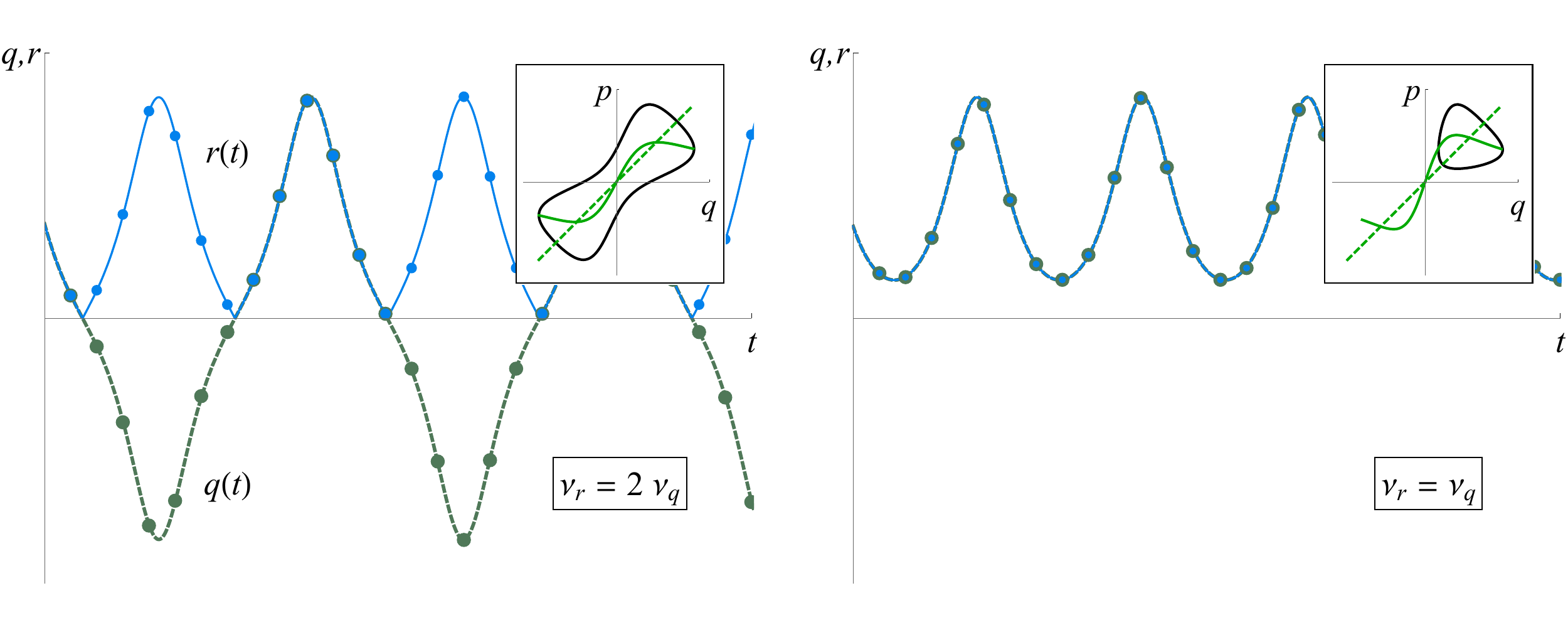}
\vspace{-0.25cm}
\caption{\label{fig:NuQNuRNuTheta}
    Parametrization of Cartesian (dark green dashed curve) and
    radial (blue curve) coordinates, regime II with $\pt=0$.
    Left and right plots illustrate trajectories outside and inside
    the separatrix.
    } 
\end{figure}
\newpage
\twocolumngrid

$\bullet$ For $[\Gamma_-]$, the fixed point at the origin remains
stable for $|a|<2$, while the symmetric fixed points
\[
(q,p)_*^{(2,3)} = \pm \sqrt{(2-a)/2}\,(1, 1)
\]
or 2-cycle
\[
(q,p)^{2-\mathrm{cycle}} = \pm \sqrt{(2+a)/2}\,(1,-1)
\]
are always unstable and defined for $a<2$ or $a>-2$
respectively.
When $a = 2$ ($a = -2$), the system undergoes the subcritical
pitchfork (subcritical period doubling) bifurcation, as depicted
in Fig.~\ref{fig:1DMcMGammaM}.
As a result, stable closed trajectories only exist when the fixed
point at the origin is stable, and the dynamic aperture is defined
by the separatrix associated with the symmetric fixed points
(2-cycle) for $a>0$ ($a<0$).

Furthermore, Fig.~\ref{fig:JNuOct} provides the rotation number
as a function of the action variable $\nu_\mathrm{oct}(J)$ (see
\cite{zolkin2022mcmillan,zolkin2025MCdynamics} for details).
The plots in the figure correspond to regimes I, II, and III from
left to right.
For all orbits in regimes I and III, as well as orbits that round
the origin and the figure-eight separatrix in regime II, the polar
rotation numbers are given by:
\[
\nu_r\Big|_{J_\theta=0} = 2\,\nu_\mathrm{oct},
\qquad\qquad\qquad\qquad
\nu_\theta\Big|_{J_\theta=0} = \nu_\mathrm{oct}.
\]
This behavior is similar to the previously considered case of the
linear oscillator. However, for orbits inside the separatrix in
regime II, the limiting behavior is defined as:
\[
\nu_r\Big|_{J_\theta=0} = \nu_\mathrm{oct},
\qquad\qquad
\nu_\theta\Big|_{J_\theta=0} = \left\{
\begin{array}{ll}
    0,   & a \geq 0,    \\[0.25cm]
    1/2, & a   <  0.
\end{array}
\right.
\]
Figure~\ref{fig:NuQNuRNuTheta} provides an illustration to explain
this difference.
The left and right plots show the parametrization of the coordinate
$q(t)$ and the radius $r(t)$ for trajectories outside and inside the
separatrix in regime II with $a>2$.
In the first case, the coordinate $q(t)$ can be negative, resulting
in $r=|q|$ oscillating at twice the frequency.
In the second case, $q(t)$ is strictly positive, leading to the
radial coordinate matching the Cartesian one, i.e., $r=q$.
For $a>0$, the trajectory does not round the origin in the
$(x,y)$-plane, and the motion is constrained by
$\{\theta\}_n = \{\theta\}_0$.
This causes $\nu_\theta$ to vanish.
For $a<2$, the trajectory jumps between two line segments with
$\{\theta\}_n = (-1)^n\,\{\theta\}_0$, resulting in $\nu_\theta=1/2$.

Finally, considering the limit $J_r \rightarrow 0$, we can determine
rotation number $\nu_0 \equiv \nu_\mathrm{oct}(J = 0)$ by evaluating
the Jacobian of the McMillan octupole map
\[
\J_{\m_\mathrm{oct}} = \left(
\begin{array}{ll}
\ds\frac{\pd q'}{\pd q} & \ds\frac{\pd p'}{\pd q} \\[0.25cm]
\ds\frac{\pd q'}{\pd p} & \ds\frac{\pd p'}{\pd p}
\end{array}
\right) =
\left(
\begin{array}{cc}
\ds 0 & 1 \\[0.45cm]
\ds-1 & \ds a\,\frac{1\,\mpt\,q^2}{(1\,\pmt\,q^2)^2}
\end{array}
\right)
\]
at the appropriate fixed point.
For cases I and III, we use the fixed point at the origin,
$q_*^{(1)}$, while for case II, we consider one of the two stable
symmetric fixed points, such as $q_*^{(2)}$, or the 2-cycle if
$a<-2$.
By computing the trace of the Jacobian
\[
\mathrm{Tr}\,\mathbf{J}_{\m_\mathrm{oct}}[q_*^{(1)}] = a
\qquad\mathrm{and}\qquad
\mathrm{Tr}\,\mathbf{J}_{\m_\mathrm{oct}}[q_*^{(2,3)}] = \frac{8}{a} - 2
\]
we obtain:
\[
\nu_0^{(1)} = \frac{1}{2\,\pi}\,\arccos\frac{a}{2},
\qquad
\nu_0^{(2)} =
    \left\{\begin{array}{ll}
        \mu_0^{(2)},   & a \geq 2  \\[0.25cm]
        1-\mu_0^{(2)}, & a \leq-2
    \end{array}\right.
\]
where $\ds\mu_0^{(2)} =
        \frac{1}{2\,\pi}\,\arccos\left(
            \frac{4}{|a|}-1
        \right)$.

\begin{figure*}[t!]
\includegraphics[width=\linewidth]{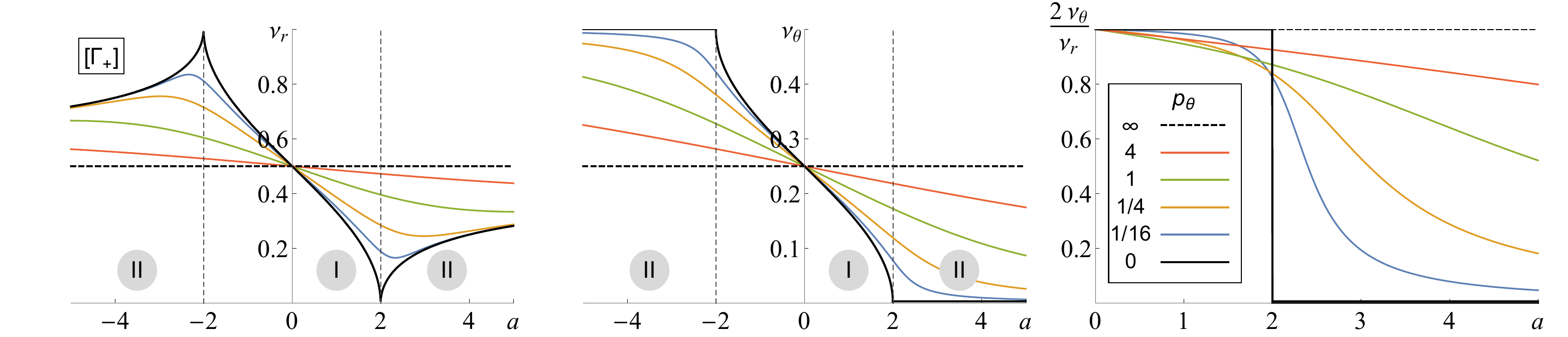}
\includegraphics[width=\linewidth]{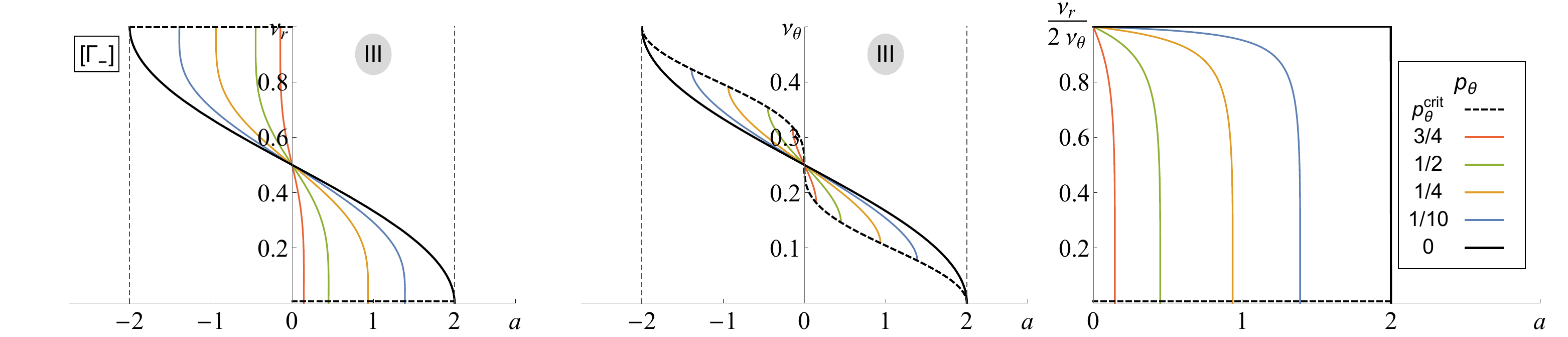}
\caption{\label{fig:Jr0Gamma}
    The radial and angular rotation numbers as functions of map
    parameter $a$ for $J_r = 0$.
    The top row corresponds to the case $[\Gamma_+]$, while the
    bottom row represents $[\Gamma_-]$.
    Each column of the figure displays a different quantity:
    $\nu_r^*$, $\nu_\theta^*$, and their ratio, respectively,
    from left to right.
    Various curves in the figure correspond to different values
    of angular momentum, as indicated in the plot legends.
    }
\end{figure*}

\subsection{Stable circular orbits, $J_r = 0$}

The last limiting case occurs when $J_r \rightarrow 0$ for a given
value of angular momentum.
This situation corresponds to a stable fixed point in the radial
degree of freedom, resulting in circular orbits $r=\const$, while
only the angular coordinate undergoes changes:
\[
    \theta' =
        \theta + \arctan \frac{\pt}{r_*^\text{st}\,\pr^\text{st}}.
\]
The resulting circle map has a constant phase advance, and the
corresponding angular rotation number is given by:
\begin{equation}
\label{math:nuTh*}
\nu_\theta^* \equiv
\nu_\theta\Big|_{J_r=0} =
    \frac{\Delta_\theta^*}{2\,\pi},
\end{equation}
with
\[
\Delta_\theta^*\,= 
    \arctan\left(
        \frac{\pt}{r_*^\text{st}\,\pr^\text{st}}
    \right) +
    \pi\,\mathrm{sgn}[\pt]\,\mathrm{H}[-a].
\]
The radial rotation number can be determined from the Jacobian
$\J_{\m_r^{\pmt}}[r_*,p_*]$
\[
\left(
\begin{array}{cc}
\ds-\frac{\pt^2}{r_*^4} &
\ds-\frac{p_*}{r_*} - \frac{p_*\,\pt^2}{r_*^5} -
2\left( \frac{2}{a}\right)^2(1-r_*^4)\,\frac{\pt^2p_*^3}{r_*^7}
\\[0.35cm]
\ds \frac{p_*}{r_*}     &
\ds-\frac{\pt^2}{r_*^4} +
2\left( \frac{2}{a}\right)^2(1-r_*^4)\,\frac{p_*^4}{r_*^4}
\end{array}
\right),
\]
by evaluating the trace for the circular orbit:
\[
\mathrm{Tr}\,\mathbf{J}_{\m_r}[r_*,p_*] = 2\,
    \frac{(2/a)^2 (1-r_*^4)\,p_*^4 - \pt^2}{r_*^4},
\]
which gives
\begin{equation}
\label{math:nuR*}
\nu_r^* \equiv
\nu_r\Big|_{J_r=0} =
    \left\{\begin{array}{ll}
        \mu_*,     & a \geq 0,          \\[0.25cm]
        1 - \mu_*, & a <    0,
    \end{array}\right.,
\end{equation}
where
\[
\mu_* = \frac{1}{2\,\pi}\,
    \arccos \frac{\mathrm{Tr}\,\mathbf{J}_{\m_r}[r_*,p_*]}{2}.
\]
In the case of $[\Gamma_+]$, equations (\ref{math:nuR*}) and
(\ref{math:nuTh*}) exhibit two important limits discussed in the
previous subsections.
On one side, as angular momentum approaches zero, we have:
\[
\lim_{J_\theta\rightarrow 0}\nu_r^* = \left\{
\begin{array}{ll}
\ds 2\,\nu_0^{(1)},     & |a|   <  2,     \\[0.25cm]
\ds \nu_0^{(2)},        & |a| \geq 2,
\end{array}
\right.
\]
and
\[
\lim_{J_\theta\rightarrow 0}\nu_\theta^* = \left\{
\begin{array}{ll}
\ds \nu_0^{(1)},        & |a|   <  2,     \\[0.4cm]
\ds 0,                  & |a| \geq 2.
\end{array}
\right.
\]
On the other side, for large amplitudes resulting from a large
value of $\pt$, we have:
\[
\lim_{J_\theta\rightarrow\infty}\nu_r^*      = \frac{1}{2},
\qquad\qquad\qquad\qquad
\lim_{J_\theta\rightarrow\infty}\nu_\theta^* = \frac{1}{4}.
\]

\begin{figure*}[t!]
\includegraphics[width=\linewidth]{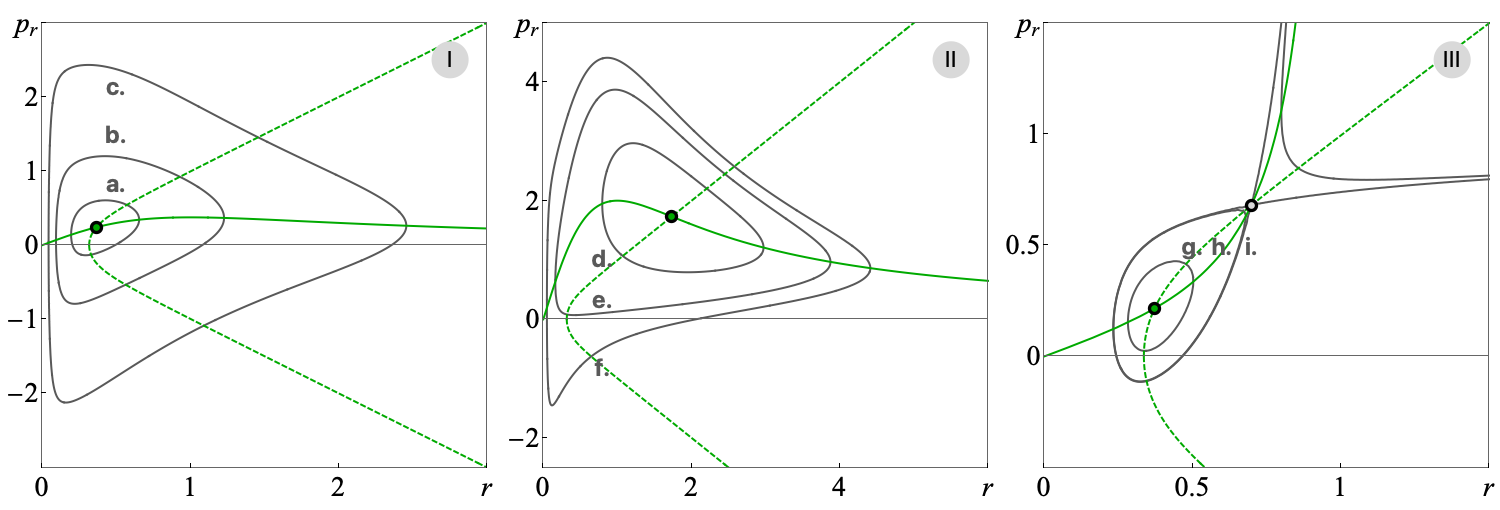}
\caption{\label{fig:StudyCases}
    Level sets of the radial invariant $\K_r[\pr,r]$ for specific
    examples labeled (a.) to (i.).
    Each plot corresponds to one of the three regimes, I to III,
    respectively.
    In case III, level sets (h.) and (i.) are chosen to be very close
    to the separatrix and each other.
    Refer to Table~\ref{tab:StudyCases} for the exact values of the
    map parameter and invariants for each example.
    }
\end{figure*}

\noindent
The top row of Fig.~\ref{fig:Jr0Gamma} illustrates the behavior of
the radial and angular rotation numbers, as well as their ratio, as
functions of $a$ for $J_r=0$ and various values of $\pt$.
As observed, while $\nu_\theta^*$ is a monotonic function of $a$,
$\nu_r^*$ is monotonic only for $|a|<2$ (regime I).
In regime II, when $a>2$ (or $<2$), $\nu_r^*$ decreases (increases)
until it reaches its minimum (maximum), and then monotonically tends
to $1/2$.
The last plot shows the ratio $2\,\nu_\theta^*/\nu_r^*$.
It is worth noting that for the first regime (I), we have
$2\,\nu_\theta^* \approx \nu_r^*$ for all values of $\pt$, while in
the second regime (II), the ratio can be approximately zero for small
values of angular momentum or large values of $a$.

In the case $[\Gamma_-]$ (regime III), the first limiting situation
is the same as for case I:
\[
\lim_{J_\theta\rightarrow 0}\nu_r^*      = 2\,\nu_0^{(1)},
\qquad\qquad\qquad
\lim_{J_\theta\rightarrow 0}\nu_\theta^* = \nu_0^{(1)}.
\]
However, now we are not interested in the limit
$J_\theta\rightarrow\infty$ since the motion is bounded only for
$\pt < \pt^\mathrm{crit}$.
Instead, we consider the limit $J_\theta\rightarrow|\pt^\mathrm{crit}|$.
The radial rotation number can be expressed using Heaviside function
and $\mu_* = 0$:
\[
\nu_r^\mathrm{crit} \equiv
\lim_{J_\theta\rightarrow J_\theta^\mathrm{crit}}\nu_r^* =
\mathrm{H}[-a] = 
\left\{\begin{array}{ll}
    0,  & a \geq 0,     \\[0.35cm]
    1,  & a  <   0.
\end{array}\right.
\]
For a given value of the map parameter, $\nu_\theta^*$ is
bounded by $\nu_0^{(1)}$ from above for $a\geq0$ (below for $a<0$)
and by
\[
\nu_\theta^\mathrm{crit} \equiv
\lim_{J_\theta\rightarrow J_\theta^\mathrm{crit}}\nu_\theta^* =
\frac{1}{2\,\pi}\,
\left\{\begin{array}{ll}
        \Delta_\theta^\mathrm{crit}, & a \geq 0,          \\[0.25cm]
        \pi - \Delta_\theta^\mathrm{crit}, & a <    0.
    \end{array}\right.,
\]
where
\[
    \Delta_\theta^\mathrm{crit} = 
    \arctan\frac{1}{\sqrt{(\pt^\mathrm{crit})^{-2/3}-1}} =
    \arctan\sqrt{\left( \frac{2}{a}\right)^{2/3}-1}.
\]
These limits are illustrated by black solid and dashed curves in
the bottom row of Fig.~\ref{fig:Jr0Gamma}.

\section{\label{sec:Regimes}Regimes of motion}

\subsection{Action-angle variables}

In the previous section, we examined the limiting cases of large
and small amplitudes.
Now, we need to bridge the gap and understand the dynamics for
typical intermediate values of the action variables.
To illustrate our results, we have selected three different values
of $J_r$ for each regime I to III, given a specific intermediate
value of angular momentum $\pt$.
These selected examples are labeled as (a.) to (i.).
Table~\ref{tab:StudyCases} provides a summary of the map parameter
$a$, the motion invariants $\K_r$ and $\pt$, as well as the action
variables $J_{r,\theta}$ and rotation numbers $\nu_{r,\theta}$ for
each case study.
Additionally, Fig.~\ref{fig:StudyCases} shows the corresponding
invariant level sets $\K_r$ in the radial phase space.
\begin{table*}[t!]\centering
\begin{tabular}{l lllrlllrlll}
\hline\hline                                                    \\[-0.3cm]
                & \multicolumn{7}{c}{$[\Gamma_+]$}      &       $\qquad$$\qquad$
                & \multicolumn{3}{c}{$[\Gamma_-]$}              \\[0.2cm]
                & \multicolumn{3}{c}{I}                 &       $\qquad$
                & \multicolumn{3}{c}{II}                &
                & \multicolumn{3}{c}{III}                       \\[0.2cm]
                & \multicolumn{3}{c}{$a=3/2$, $\pt =-0.1$}      &
                & \multicolumn{3}{c}{$a=8  $, $\pt = 0.1$}      &
                & \multicolumn{3}{c}{$a=1  $, $\pt = 0.5\,\pt^\mathrm{crit}
                                                   = 0.1125$}   \\[0.2cm]
                &$\quad$(a.)&$\quad$(b.)&$\quad$(c.)    &
                &$\quad$(d.)&$\quad$(e.)&$\quad$(f.)    &
                &$\quad$(g.)&$\quad$(h.)&$\quad$(i.)            \\[-0.3cm]
                                                                \\\hline
                                                                \\[-0.3cm]
$\, J_r\qquad\qquad$
                & 0.04      & 0.25      & 1             &
                & 0.5       & 1.3695    & 2             &
                & 0.01      & 0.033692  & 0.0344042             \\[0.2cm]
$\, \K_r$       & 0.28228   & 1.17187   & 5.54626       &
                &-5.54758   & 0         & 4.2368        &
                & 0.217948  & 0.2746    & 0.275683              \\[0.2cm]
$\, \nu_r$      & 0.310521  & 0.375294  & 0.440763      &
                & 0.310913  & 0.244355  & 0.310346      &
                & 0.269487  & 0.166497  & 0.0785715             \\[0.2cm]
$\, \nu_\theta$ & 0.148603  & 0.184318  & 0.219254$\qquad$&
                & 0.007675  & 0.062641  & 0.146941      &
                & 0.144787  & 0.102572  & 0.0680104     $\,$    \\[0.1cm]
\hline\hline
\end{tabular}
    \caption{\label{tab:StudyCases}
    Angular and radial invariants, rotation numbers, and action
    variables for specific examples in regimes I -- III.
    The angular action variable is given by $J_\theta = |\pt|$.
    }
\end{table*}

\begin{figure}[h!]\centering
\includegraphics[width=0.71\linewidth]{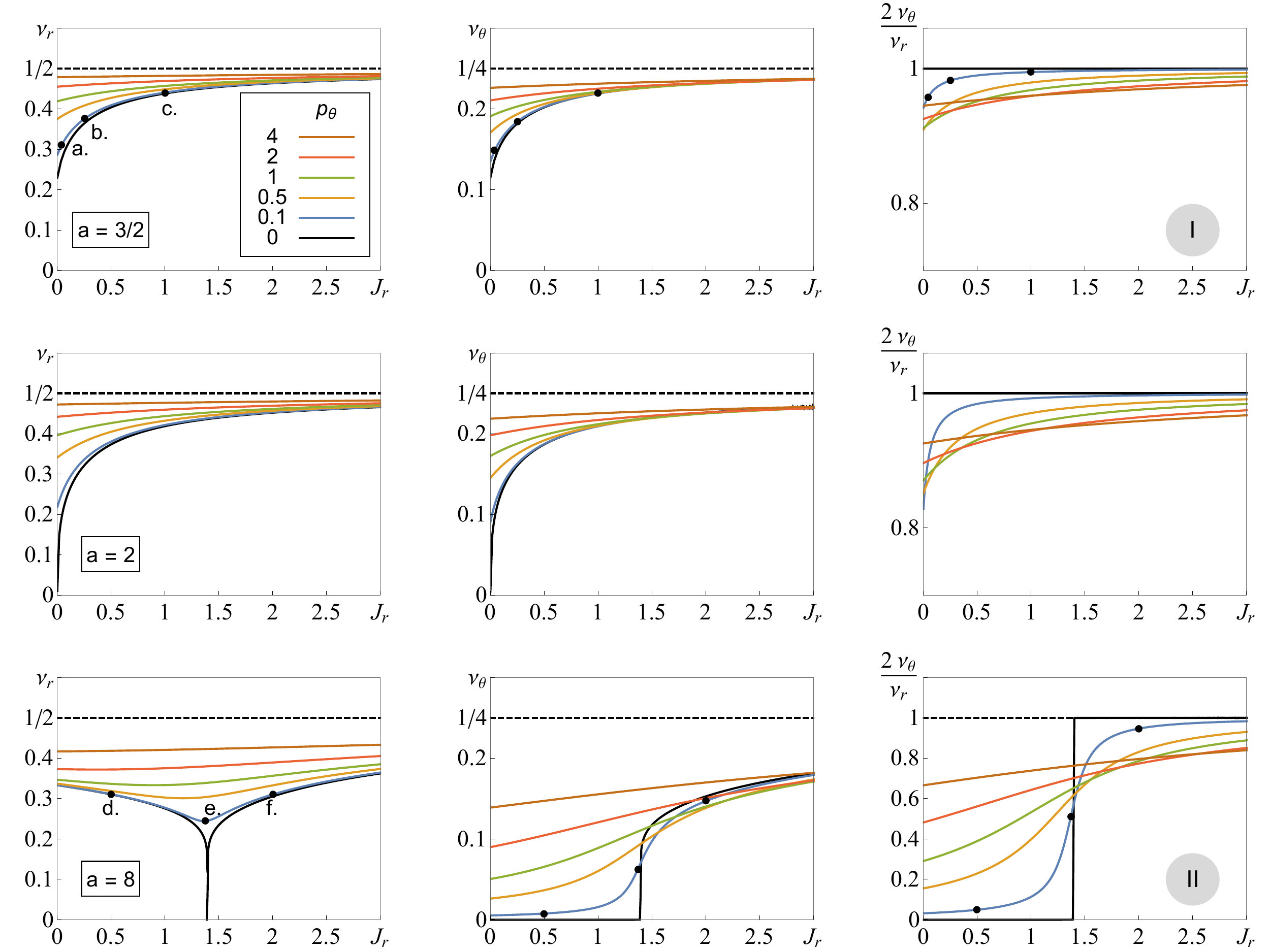}
\caption{\label{fig:ActionsP}
    Radial and angular rotation numbers $\nu_{r,\theta}$ and their
    ratio as functions of radial action $J_r$ for the $[\Gamma_+]$
    configuration.
    Different values of the angular momentum $\pt$ are depicted
    with colors according to the legend.
    Dashed lines represent the limit of large radial amplitudes,
    where $J_{r,\theta} \rightarrow \infty$.
    Rows corresponds to different values of $a=3/2,2,8$.
    Black dots indicate points for further analysis, with (a.) to
    (c.) representing regime I and (d.) to (f.) representing
    regime II.
    }
\end{figure}

Before delving into the detailed analysis of each case study,
let's first examine how rotation numbers depend on actions.
Figure~\ref{fig:ActionsP} illustrates the behavior of $\nu_{r}$
and $\nu_{\theta}$ as functions of the radial action $J_r$ for
the $[\Gamma_+]$ configuration.
Each plot displays curves corresponding to different values of
angular momentum $\pt$.
The top row represents the typical situation for regime I
($|a|<2$), where both rotation numbers monotonically increase with
respect to the action variables.
For an ensemble of particles within a finite radius, spreads of
rotation numbers $\Delta\nu_{r,\theta}$ are limited.
When $J_\theta$ is fixed, they satisfy inequalities:
\[
J_\theta = \const:\,\,
    \Delta\nu_r <
    \frac{1}{2} - \nu_r^* <
    \frac{1}{2} - \frac{\arccos(a/2)}{\pi} <
    \frac{1}{2}
\]
and
\[
\Delta\nu_\theta <
    \frac{1}{4} - \nu_\theta^* <
    \frac{1}{4} - \frac{\arccos(a/2)}{2\,\pi} <
    \frac{1}{4}.
\]
Similarly, when $J_r$ is fixed, we have:
\[
\begin{array}{l}
\ds J_r = \const:\,\,
    \Delta\nu_r <
    \frac{1}{2} - 2\,\nu_\mathrm{oct} <
    \frac{1}{2} - \frac{\arccos(a/2)}{\pi} <
    \frac{1}{2},                        \\[0.45cm]
\ds \Delta\nu_\theta <
    \frac{1}{4} - \nu_\mathrm{oct} <
    \frac{1}{4} - \frac{\arccos(a/2)}{2\,\pi} <
    \frac{1}{4}.
\end{array}
\]
The largest spread occurs for the curve with zero angular momentum,
which is smaller than $1/2$ for the radial degree of freedom and
$1/4$ for the angular degree of freedom.
The ratio of rotation numbers remains locked at $1/2$ for both
limits $|\pt|=0$ and $|\pt|=\infty$, but it slightly varies for
intermediate situations.

The middle row corresponds to the case $a=2$, where the system
undergoes a supercritical pitchfork bifurcation.
In this case, the radial spread reaches its maximum value of $1/2$
for stable circular orbits and orbits with $\pt = 0$, while the
angular spread reaches its maximum value of $1/4$.

Finally, the last row corresponds to regime II (case $|a| > 2$),
where we specifically chose $a = 8$.
Recall that for $\pt = 0$, we have trajectories both inside and
outside the figure-eight separatrix.
As a consequence, we observe that the ratio of rotation numbers
is given by the Heaviside step function:
\[
\left.\frac{\nu_\theta}{\nu_r} \right|_{\pt=0} =
    \frac{1}{2}\,\mathrm{H}(J_r-J_\mathrm{sep}).
\]
When $\pt \neq 0$, the bifurcation is unexpectedly removed, and
the dependencies $\nu_{r,\theta }(J_r,J_\theta )$ become more
intricate.
The radial rotation number is monotonic only with respect to
$|\pt|$, while the angular rotation number and $\nu_\theta/\nu_r$
are monotonic only as functions of $J_r$.
In the case $|a| \leq 2$, the resonant condition
\[
    1/4 < \nu_\theta/\nu_r < 1/2
\]
holds for all trajectories.
However, in case II, we observe that for $J_r<J_\mathrm{sep}$,
the ratio of rotation numbers is monotonic with respect to both
actions, and for a fixed value of $J_r$, we have
\[
    0 < \nu_\theta/\nu_r < 1/2.
\]
The system now crosses the coupling resonance $\nu_\theta/\nu_r=1/4$,
resulting in two different types of motion: one with
$J_r > J_\mathrm{sep}$ similar to case I, and a new type with
$J_r < J_\mathrm{sep}$ for small to intermediate values of $\pt$.
Examples (d.) and (f.) were chosen to illustrate the difference,
while case study (e.) is close to the boundary between different
modes of oscillation, with
\[
\nu_\theta/\nu_r \approx 1/4.
\]

\begin{figure}[h!]\centering
\includegraphics[width=0.71\linewidth]{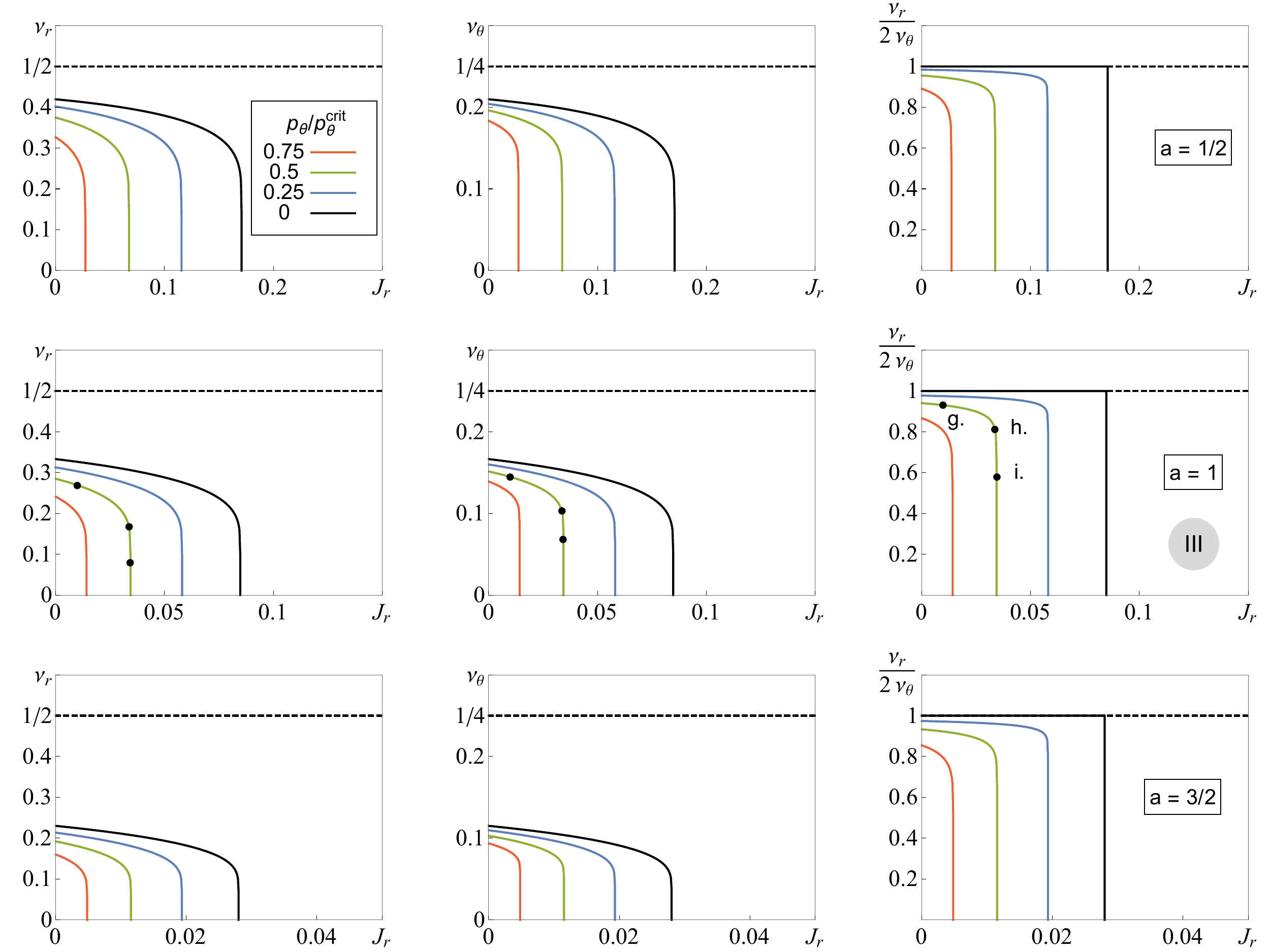}
\caption{\label{fig:ActionsM}
    Same as Fig.~\ref{fig:ActionsP} but for the $[\Gamma_-]$
    configuration and $\pt$ measured in units of $\pt^\mathrm{ctir}$.
    Rows corresponds to different values of $a=1/2,1,3/2$.
    Black dots indicate points for further analysis, with (g.) to
    (i.) representing regime II.
    } 
\end{figure}

Figure~\ref{fig:ActionsM} is similar to Figure~\ref{fig:ActionsP},
but for configuration $[\Gamma_-]$ and regime III.
Again, different rows correspond to different settings of the
parameter $a$, which is equal to 1/2, 1, and 3/2.
Case studies (g.) -- (i.) are chosen for $a=1$ and
$\pt = 0.5\,\pt^\mathrm{crit}$ (green curves in the middle row).
In this case, the value of angular momentum is measured in units
of $\pt^\mathrm{crit} \approx 0.225$, and the ratio of rotation
numbers is inverse compared to the previous examples in Figure
\ref{fig:ActionsP}.
All functions are now monotonic with respect to both actions.
Assuming that the entire radial phase space contained within the
separatrix is occupied by particles, the spread of frequencies
for fixed $J_r$ is bounded between 0 and the frequencies defined
by the 1D octupole limit:
\[
\begin{array}{l}
\ds J_r = \const:\quad
    \Delta\nu_r <
    2\,\nu_\mathrm{oct} <
    \frac{\arccos(a/2)}{\pi} <
    \frac{1}{2},            \\[0.45cm]
\ds \Delta\nu_\theta <
    \nu_\mathrm{oct} <
    \frac{\arccos(a/2)}{2\,\pi} <
    \frac{1}{4}.
\end{array}
\]
For fixed $\pt$, the upper limit is given by circular orbits with
$J_r = 0$:
\[
\begin{array}{l}
\ds J_\theta = \const:\quad
    \Delta\nu_r <
    \nu_r^* <
    \frac{\arccos(a/2)}{\pi} <
    \frac{1}{2},            \\[0.45cm]
\ds \Delta\nu_\theta <
    \nu_\theta^* <
    \frac{\arccos(a/2)}{2\,\pi} <
    \frac{1}{4}.
\end{array}
\]
The largest variation of either of the rotation numbers as functions
of $J_r$ occurs close to the separatrix.
For example, case studies (h.) and (i.) are chosen to have very close
values of radial actions, approximately $J_\mathrm{sep}$, while their
rotation numbers vary by a significant amount around $0.1$.

\newpage

\onecolumngrid

\newpage

\twocolumngrid

\subsection{Cartesian frequencies}

So far, we have focused on describing our system in polar
coordinates.
However, it is also important to consider the frequencies observed
in Cartesian coordinates.
In Cartesian coordinates, the conventional horizontal and vertical
tunes do not exist due to the axial symmetry of the system.
The spectra of oscillations in both planes are identical.
Each plane exhibits two families of overtones:
\[
\begin{array}{llllll}
(\nu_r - \nu_\theta) + n\,\nu_r:  &\,
   \nu_r - \nu_\theta,  &\,
2\,\nu_r - \nu_\theta,  &\,
3\,\nu_r - \nu_\theta,  &\,
\cdots,                 \\[0.25cm]
\nu_\theta + n\,\nu_r:  &\,
\nu_\theta,             &\,
\nu_\theta +    \nu_r,  &\,
\nu_\theta + 2\,\nu_r,  &\,
\cdots,
\end{array}
\]
for $n=0,1,2,\ldots$.
These overtones are based on the two fundamental tunes:
\[
\nu_r - \nu_\theta
\qquad\qquad\qquad\mathrm{and}\qquad\qquad\qquad
\nu_\theta.
\]
It is important to note that Cartesian coordinates are obtained by
discretizing a multiplication of two periodic functions:
\[
    x = r(t)\times\cos\theta(t)
    \qquad\quad\mathrm{and}\quad\qquad
    y = r(t)\times\sin\theta(t).
\]
This results in amplitude modulated signals.
Therefore, the fundamental tunes play the role of the sum and
difference of the "carrier" and "modulating" frequencies, $\nu_1$
and $\nu_2$ respectively:
\[
\begin{array}{lll}
\nu_\Sigma = \nu_1 + \nu_2,                 &
\qquad\qquad\qquad\qquad        & \ds
\nu_1 = \frac{\nu_\Sigma + \nu_\Delta}{2},  \\[0.45cm]
\nu_\Delta = \nu_1 - \nu_2,                 &
\qquad\qquad\qquad\qquad        & \ds
\nu_2 = \frac{\nu_\Sigma - \nu_\Delta}{2}.
\end{array}
\]
For configuration $[\Gamma_+]$, where $2\,\nu_\theta < \nu_r$, we
define:
\[
\begin{array}{lll}
\nu_\Sigma = \nu_r - \nu_\theta,            &
\qquad\qquad\qquad\qquad        & \ds
\nu_1 = \frac{\nu_r}{2},                    \\[0.45cm]
\nu_\Delta = \nu_\theta,                    &
\qquad\qquad\qquad\qquad        & \ds
\nu_2 = \frac{\nu_r}{2} - \nu_\theta.
\end{array}
\]
While for configuration $[\Gamma_-]$, where $2\,\nu_\theta > \nu_r$,
we will use:
\[
\begin{array}{lll}
\nu_\Sigma = \nu_\theta,                    &
\qquad\qquad\qquad\qquad        & \ds
\nu_1 = \frac{\nu_r}{2},                    \\[0.45cm]
\nu_\Delta = \nu_r - \nu_\theta,            &
\qquad\qquad\qquad\qquad        & \ds
\nu_2 =  \nu_\theta - \frac{\nu_r}{2}.
\end{array}
\]

\subsection{Case studies}

In this section, we analyze the general dynamics of all three
regimes, I to III.
We present the results in three collated figures: Figs.
\ref{fig:Study_I}, \ref{fig:Study_II}, and \ref{fig:Study_III}.
Each figure consists of different columns corresponding to
specific examples (a.) -- (i), and various rows dedicated to
different sets of variables.
The first row of each figure shows the iterations and the
corresponding parametrization of all polar coordinates, similar
to Figs.~\ref{fig:RPrParametrization} and
\ref{fig:ThetaParametrization}.
We represent the radial coordinate $r$ in orange, the radial
momentum $\pr$ in blue, and the angular coordinate $\theta$ in
purple.
The second row of figures illustrates the long-term behavior of
the parametrizations for the Cartesian coordinates $x$ and $y$.
The third and fourth rows depict the projection of the orbit
onto the Cartesian degrees of freedom: the physical space
$(x,y)$-plane and the phase space plane $(x,p_x)$, respectively.
The green points correspond to the iterations of the map
(approximately $10^4$), while the solid curve represents the
continuous parametrization (approximately $10$ radial oscillations).
To aid readers, the first two radial oscillations on the
continuous curve are colored in black, while the rest are shown
in red.
Finally, the bottom row presents the absolute value of the discrete
Fourier transform for the Cartesian coordinates obtained through the
iteration of the map.
The plot is shown on a logarithmic scale.
These visualizations will provide us with a comprehensive
understanding of the system's dynamics and facilitate the analysis
of different cases within each regime.

\subsubsection{Regime $\mathrm{I}$}

\begin{figure}[h!]\centering
\includegraphics[width=0.96\linewidth]{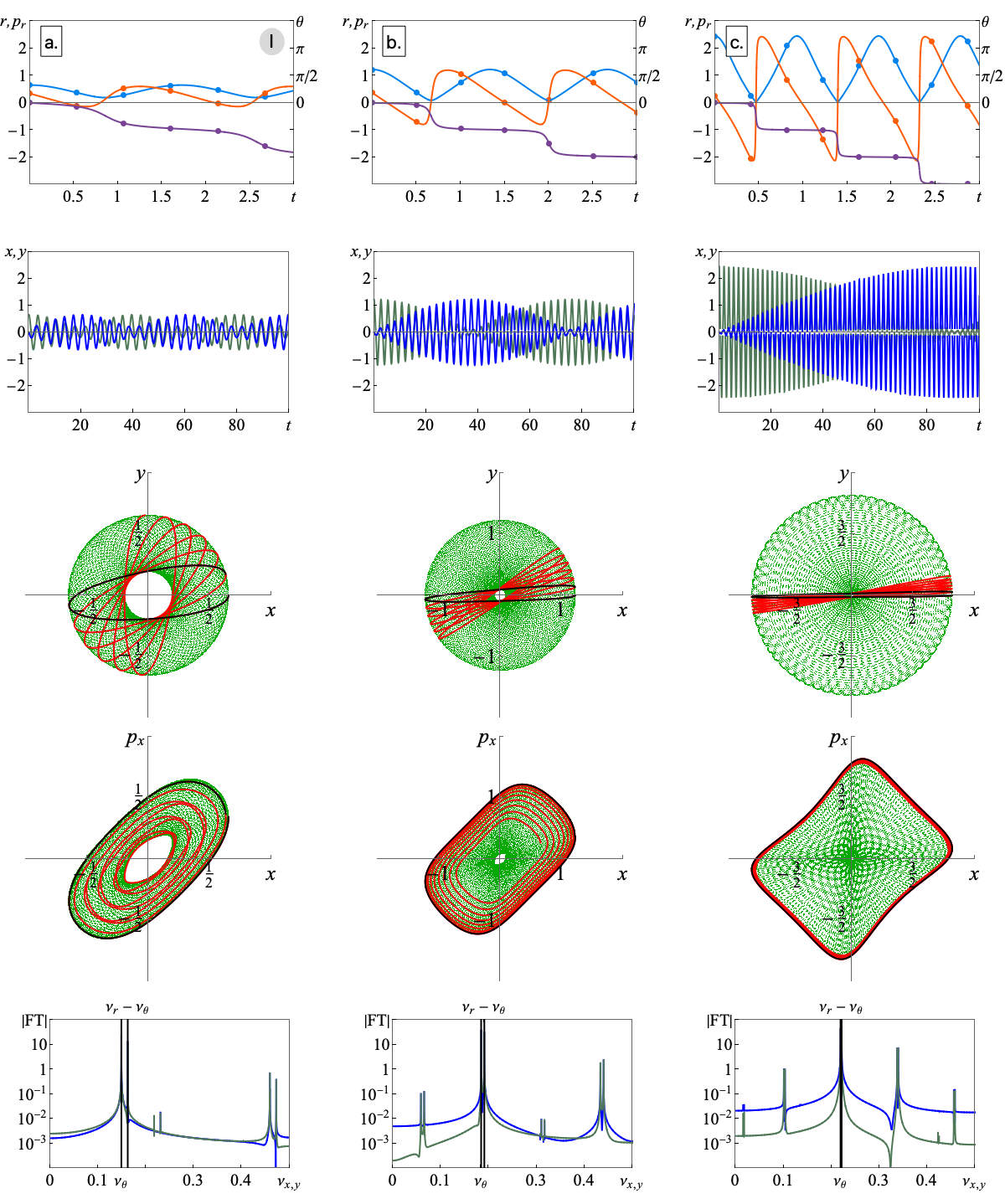}
\vspace{-0.5cm}
\caption{\label{fig:Study_I}
    Case studies (a.) -- (c.), configuration $[\Gamma_+]$ regime I.
    Top row. Short term behavior in polar coordinates.
    Solid curves show parametrization of the radius $r$ (orange),
    radial momentum $\pr$ (blue) and angular coordinate $\theta$
    (purple).
    Discretization $t=n\,\T'$ is equivalent to the iterates of the
    map.
    Second row. Long term behavior of the parametrization of the map
    in Cartesian coordinates, $x,y$.
    Third and fourth rows. Projection of orbit onto $(x,y)$- and
    $(x,p_x)$-planes. Green dots show $10^4$ iterations under the
    map, while short-term behavior of its parametrization is shown
    in black (first two radial oscillations) and red (next 10
    oscillations).
    Bottom row. Absolute value of the discrete Fourier transform
    for Cartesian iterates of the map, $\nu_{x,y}$.
    } 
\end{figure}

Let's begin with regime I, where we examine the long-term behavior
of the map's parametrization in Cartesian coordinates, $x,y(t)$.
In all cases (a.) through (c.), we observe wave packets with a
distinct beat pattern (second row in Fig.~\ref{fig:Study_I}).
The presence of beats indicates that the sum and difference modes
have nearly the same frequencies:
\[
    \nu_\Sigma \approx \nu_\Delta
    \qquad\qquad\qquad(\mathrm{or}\,\,
    \nu_r \approx 2\,\nu_\theta
    ).
\]
This implies that $\nu_1$ is much larger than $\nu_2$, resulting
in:
\[
    \nu_1 \approx \nu_{\Sigma,\Delta}
    \qquad\qquad\mathrm{and}\qquad\qquad
    \nu_2 \approx 0,
\]
as shown in the bottom row of Fig.~\ref{fig:Study_I} (Cartesian
spectra).
The presence of a beat pattern suggests that the coupling between
the Cartesian coordinates is weak.
As depicted in the third row of Fig.~\ref{fig:Study_I}, the
oscillations in the $(x,y)$-plane resemble elliptic orbits with a
gradual rotation of the major axis such that a precession angle
less than $\pi/2$.
The transfer of energy between the Cartesian degrees of freedom
occurs gradually in distinct stages, which we refer to as a
``weak coupling mode.''

\subsubsection{Regime $\mathrm{II}$}

\begin{figure}[h!]\centering
\includegraphics[width=\linewidth]{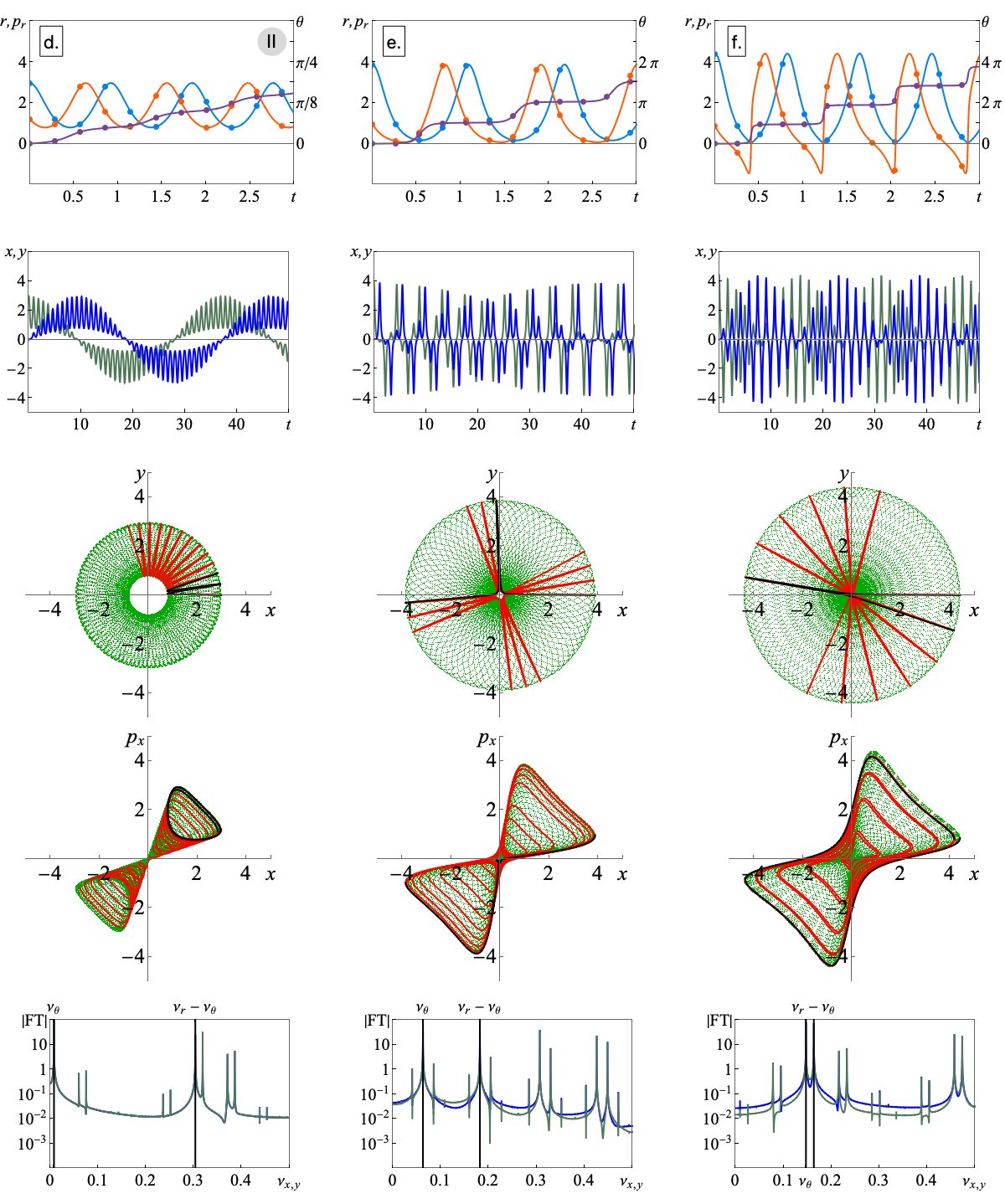}
\vspace{0.5cm}
\caption{\label{fig:Study_II}
    Same as in Fig.~\ref{fig:Study_I}, but for regime II,
    case studies (d.) -- (f.).
    } 
\end{figure}

Next, let's consider regime II.
Looking at the Cartesian phase space (fourth row in Fig.
\ref{fig:Study_II}), we can observe similarities to motion for the
1D octupole map with $\pt=0$.
In case (f.), the trajectory rounds the origin, in case (e.), we
observe a trajectory similar to a figure-eight pattern, and for
small radial action, case (d.), the trajectory over the course of
one radial oscillation resembles trajectories inside the
figure-eight separatrix.

\newpage

\onecolumngrid

\newpage

\twocolumngrid

The third row in Fig.~\ref{fig:Study_II} shows the same orbits
projected onto the configuration space.
We can see that the main difference from 1D dynamics is the
presence of orbital precession.
Comparing these plots to the second row, which represents the
long-term behavior of $x$ and $y$, we notice that definite beat
patterns are present only for the case (f.) with large radial
amplitude.
This situation somewhat resembles the larger amplitudes observed
in regime I, where the particle approximately rounds the origin in
the $(x,y)$-plane per two radial oscillations.

In contrast, for case (d.) where $\nu_r > \nu_\theta \approx 0$,
we need many radial periods to complete one orbit around the origin.
The horizontal and vertical oscillations follow the angular variable
with a high-frequency jittery motion caused by the radial
oscillations.
The absence of a clear beat pattern indicates that this is a
different mode of oscillations with $\nu_\Sigma \gg \nu_\Delta$.
We will refer to this mode as the ``strong coupling mode.''

Case (e.) separates the two modes of oscillations.
Although it is not a true separatrix, it is in the vicinity of a
closed orbit with a resonant condition of
$4\,\nu_\theta=\nu_r$, resembling a figure-eight separatrix
(as discussed in the next section).
Specifically, for the chosen examples, we have the following
relations between the sets of frequencies:
\[
\begin{array}{lcc}
(d.)& \qquad\ds
\nu_\theta,\,\frac{\nu_\theta}{\nu_r} \approx 0,
    & \qquad\qquad\ds
    \nu_\Sigma \gg \nu_\Delta \approx 0,        \\[0.35cm]
(e.)& \qquad\ds
    0 < \frac{\nu_\theta}{\nu_r} < \frac{1}{2},
    & \qquad\qquad\ds
    \nu_\Sigma > \nu_\Delta,                    \\[0.35cm]
(f.)& \qquad\ds
    \frac{\nu_\theta}{\nu_r} \approx \frac{1}{2},
    & \qquad\qquad\ds
    \nu_\Sigma \approx \nu_\Delta,
\end{array}
\]
or in terms of $\nu_{1,2}$:
\[
\begin{array}{lc}
(d.)& \qquad\qquad\ds
\nu_1 \approx \nu_2 \approx \nu_{\Sigma}/2,         \\[0.35cm]
(e.)& \qquad\qquad\ds
\nu_1 > \nu_2 \approx 0,                            \\[0.35cm]
(f.)& \qquad\qquad\ds
(\nu_1 \approx \nu_{\Sigma,\Delta}) \gg (\nu_2 \approx 0).
\end{array}
\]

\subsubsection{Regime $\mathrm{III}$}

The last case we consider is configuration $[\Gamma_-]$ in regime
III.
Examining the relation between sets of frequencies
\[
\begin{array}{lcc}
(g.)& \qquad\ds \frac{\nu_\theta}{\nu_r} \approx \frac{1}{2},
    & \qquad\qquad\ds \nu_\Sigma \approx \nu_\Delta,        \\[0.35cm]
(h.)& \qquad\ds \frac{1}{2} < \frac{\nu_\theta}{\nu_r} < 1,
    & \qquad\qquad\ds \nu_\Sigma > \nu_\Delta,              \\[0.35cm]
(i.)& \qquad\ds \frac{\nu_\theta}{\nu_r}  \approx 1,
    & \qquad\qquad\ds \nu_\Sigma \gg \nu_\Delta \approx 0,
\end{array}
\]

\begin{figure}[h!]\centering
\includegraphics[width=\linewidth]{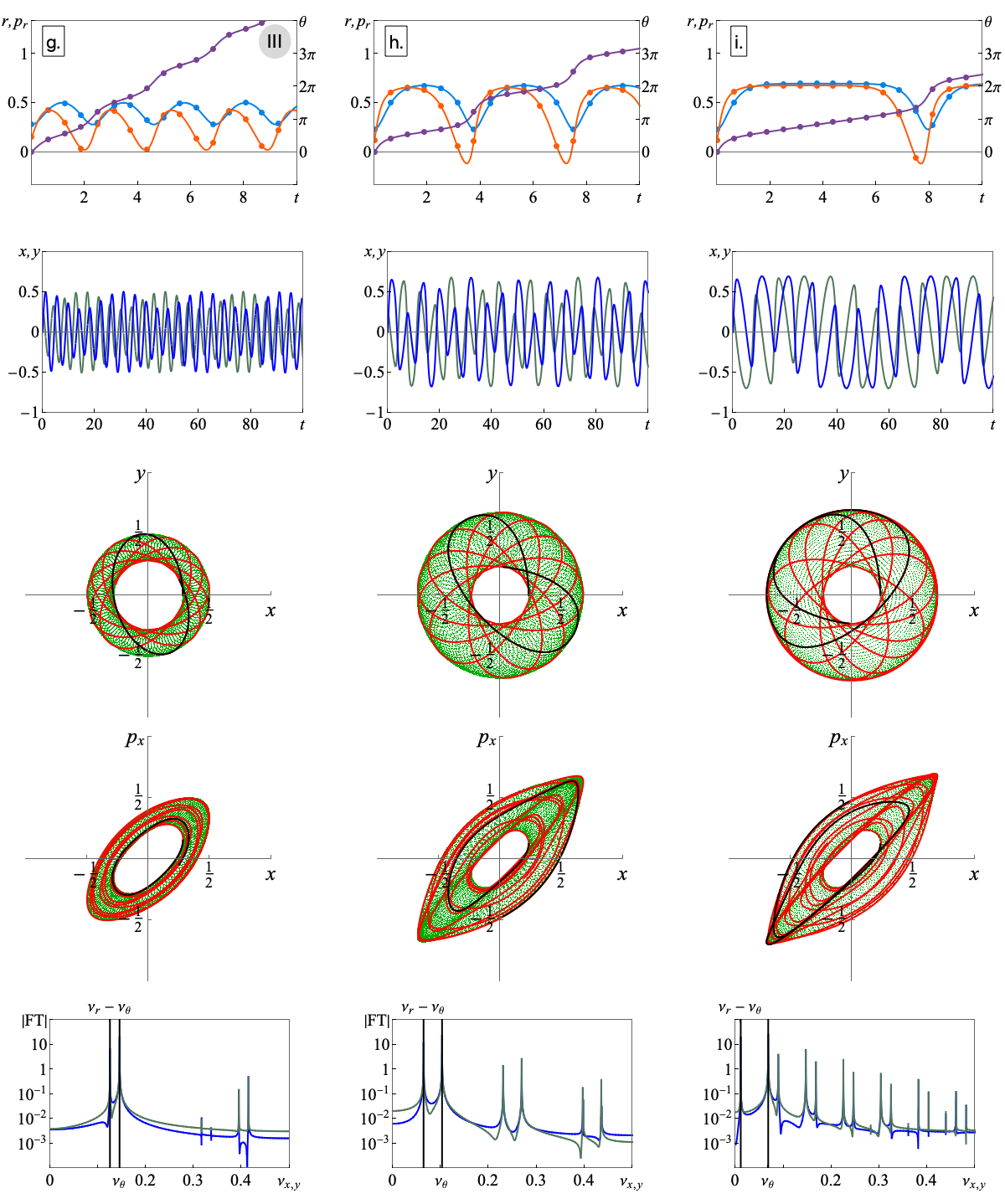}
\caption{\label{fig:Study_III}
    Same as in Fig.~\ref{fig:Study_I}, but for configuration
    $[\Gamma_-]$ regime III, case studies (g.) -- (i.).
    } 
\end{figure}

\newpage
\noindent
and
\[
\begin{array}{lc}
(g.)& \qquad\qquad\ds (\nu_1 \approx \nu_{\Sigma,\Delta})
        \gg (\nu_2 \approx 0),                              \\[0.35cm]
(h.)& \qquad\qquad\ds \nu_1 > \nu_2 \approx 0,              \\[0.35cm]
(i.)& \qquad\qquad\ds \nu_1 \approx \nu_2 \approx 0,
\end{array}
\]
we observe a situation that is the reverse of regime II. 
For small radial actions (g.), we have weak coupling with clear
beats and $\nu_\Sigma \approx \nu_\Delta$.
However, for trajectories close to the separatrix (h.,i.),
$\nu_\Sigma$ and $\nu_\Delta$ separate, and the beats change to a
jittery oscillations.
In this case, the motion of $x$ and $y$ follows the slow mode of
$\nu_r - \nu_\theta$, while the jitters are caused by the high
frequency $\nu_\theta\approx\nu_r$.
Refer to the second row of Fig.~\ref{fig:Study_III} for the
long-term behavior of the Cartesian coordinates and the bottom row
for their spectra.
Once again, the two modes of oscillations are separated by small
and large amplitudes.
However, it is important to note that the strong coupling mode only
occurs for particles very close to the separatrix and occupies a
tiny fraction of the phase space.

\subsection{Closed trajectories}

\begin{figure}[h!]\centering
\includegraphics[width=0.97\linewidth]{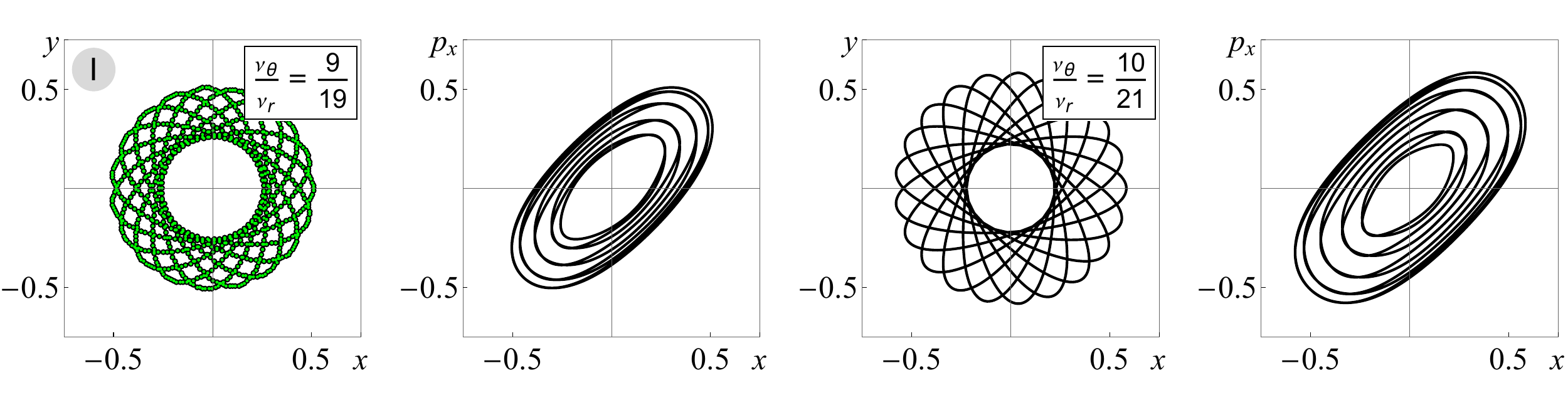}
\includegraphics[width=0.97\linewidth]{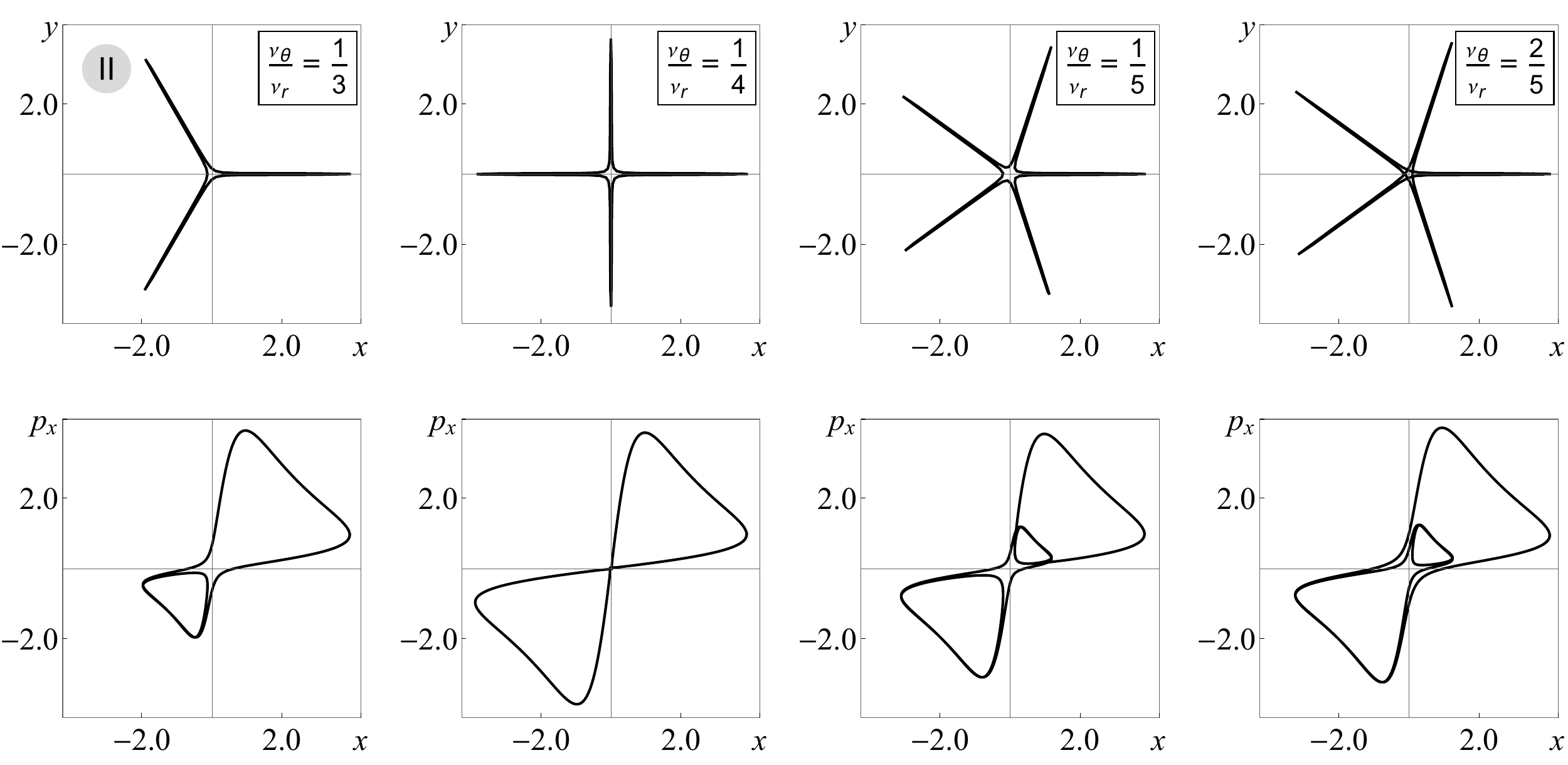}
\includegraphics[width=0.97\linewidth]{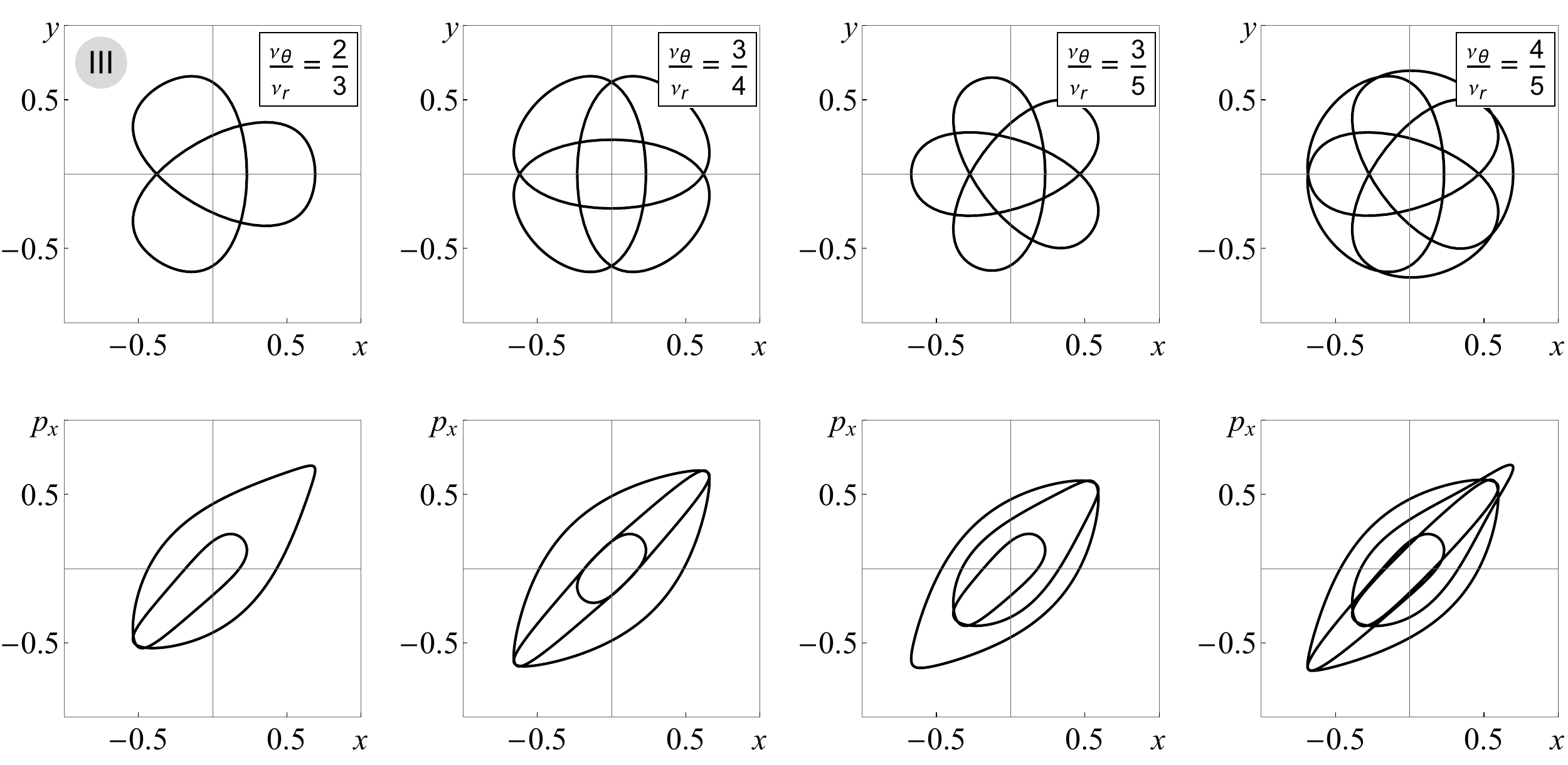}
\caption{\label{fig:ResTraj}
    The projection of closed 4D trajectories onto the $(x,y)$ and
    $(x,p_x)$ planes.
    The green dots represent $10^3$ iterations of the map, while
    the black curve behind them represents the parametrization of
    the trajectory (example with $\nu_\theta/\nu_r =9/19$). 
    The top row of figures represents regime I, while the next two
    rows correspond to regime II.
    The last two rows depict regime III.
    The parameter $a$ and angular momentum $\pt$ have the same
    values as the ones used in the case studies.
    } 
\end{figure}

It is well known that in the general central-force problem, most
of the orbits in the $(x,y)$-plane are not closed and eventually
pass arbitrarily close to every point within the annulus.
All orbits are closed only for linear (2D isotropic harmonic
oscillator) and inverse-square (Kepler problem) laws
~\cite{arnold1990huygens,santos2007english}.
To have a closed orbit, it must satisfy the resonant condition
\[
    m\,\nu_r - n\,\nu_\theta = 0,
\]
where $m$ and $n$ are positive integers.
In Fig.~\ref{fig:ResTraj}, we present a few examples of
resonant trajectories for the same case studies in regimes II and
III.
For each example, we show the projection of the trajectory onto the
horizontal phase space and the $(x,y)$-plane.

\newpage

\onecolumngrid

\newpage

\twocolumngrid

\section{\label{sec:RoundBeams}Approximated radial invariant for
round beams}

Let us consider an accelerator lattice composed of a special
linear insert~(\ref{math:LinearInsert}) followed by an axially
symmetric thin lens:

\[
\begin{bmatrix}
    x \\[0.082cm] \dot{x} \\[0.082cm] y \\[0.082cm] \dot{y}
\end{bmatrix}' =
\begin{bmatrix}
    x \\[0.082cm] \dot{x} \\[0.082cm] y \\[0.082cm] \dot{y}
\end{bmatrix}
+
\begin{bmatrix}
    0                               \\[0.05cm]
\ds \delta \dot{R}(r)\,\cos\theta   \\[0.05cm]
    0                               \\[0.05cm]
\ds \delta \dot{R}(r)\,\sin\theta
\end{bmatrix}.
\]
Here, the radial kick $\delta\dot{R}(r)$ is assumed to be a
differentiable odd function (otherwise, the system would have
singularities at $r=0$) with $\delta\dot{R}(0) = 0$, ensuring
an equilibrium orbit at the origin for $\pt=0$, but arbitrary
otherwise.

By employing the transformation (\ref{math:CanTrans}), we can
express this map in the McMillan
form~\cite{mcmillan1971problem,turaev2002polynomial}:
\[
\begin{array}{ll}
    \ds q'   &\ds\!\!=   p_q,                             \\[0.35cm]
    \ds p_q' &\ds\!\!= - q + \delta\dot{r}(r')\,\frac{q'}{r'},
\end{array}
\]
or in polar coordinates as
\[
\begin{array}{l}
    \ds r'\,= \sqrt{\pr^2 + \frac{\pt^2}{r^2}},             \\[0.4cm]
    \ds \pr'= \,-\pr\,\frac{r}{r'} + \delta\dot{r}(r'),
\end{array}
\qquad\qquad
\begin{array}{l}
    \ds \theta'\,= \theta + \arctan\frac{\pt}{r\,\pr},         \\[0.65cm]
    \ds \pt'= \pt,
\end{array}
\]
where the new radial kick is given by
\[
\delta\dot{r}(r) =
    2\,r\,\cos\Phi + \beta\,\delta\dot{R}(r)\,\sin\Phi.
\]

Despite the separation of variables, for a general kick function
$\delta\dot{r}(r)$, the map is known to exhibit chaotic behavior
and only possesses partial integrability, with the axial symmetry
giving rise to the exact invariant of motion
\[
    \K_\theta[\pt,\theta] = \pt.
\]
However, in typical situations where $\delta\dot{R}(r)$ can be
expanded around the origin, it can be shown that the system has
an additional approximated radial invariant given by
\[
\K_r[\pr,r] \approx \mathrm{C.S.}_r -
            \frac{c}{3!}\frac{\Pi_r^2}{a} +
            \frac{\pt^2}{r^2},
\]
where the parameters $a$ and $c$ are defined as
\[
    a = 2\,\cos\Phi + \beta\,\sin\Phi\,\pd_r\delta\dot{R}(0),
    \qquad
    c = \beta\,\sin\Phi\,\pd_{rrr}\delta\dot{R}(0).
\]
Here, $\mathrm{C.S.}_r = \Sigma_r^2 - (a+2)\,\Pi_r$ represents the
radial Courant-Snyder term expressed in symmetric notations
\[
    \Sigma_r = \pr + r,
    \qquad\qquad\qquad\qquad\qquad\qquad
    \Pi_r = \pr\,r.
\]
Therefore, for small radial displacements, we can approximate the
dynamics in nonlinear round beam optics using the axially symmetric
McMillan map.

\newpage
\section{\label{sec:Summary}Summary}

In this article, we considered the transverse dynamics of a single
particle in an integrable accelerator lattice utilizing the McMillan
axially symmetric electron lens.
While McMillan e-lens has the potential to mitigate collective space
charge forces, certain fundamental aspects of this device remained
unexplored.
Thus, our primary objective was to bridge this gap and gain a
comprehensive understanding of its limitations and potential.

Additional significance of the McMillan axially symmetric map lies
in its provision of first-order approximations for the dynamics of
a general linear lattice combined with an arbitrary thin lens, where
the motion can be separated in polar coordinates.
Therefore, by deepening our comprehension of this map, we can obtain valuable insights into the behavior of round beams that are not
necessarily integrable.

We performed a classification of all feasible regimes exhibiting
stable trajectories and determined the corresponding canonical
action-angle variables.
This analysis enables us to evaluate essential quantities such as
the dynamical aperture, Poincar\'e rotation numbers, and the spread
in nonlinear tunes.
Moreover, we established a parametrization of invariant curves,
facilitating the direct determination of the map's image.

In the second part of the article, we explored the dynamics as a
function of system parameters.
Our investigation reveals three fundamentally different configurations
of the accelerator optics, leading to distinct regimes of nonlinear
oscillations.
We provide a comprehensive analysis of each regime, including the examination of limiting cases for large and small amplitudes.
Additionally, we consider the dynamics in Cartesian coordinates and
provide a description of observable variables and corresponding
spectra.

Overall, this study provides valuable insights into the transverse
dynamics of particles in an integrable accelerator lattice with the
McMillan axially symmetric electron lens.
Our findings not only deepen our understanding of this specific
device but also contribute to the broader understanding of round
beam behavior in accelerator systems.

\section{\label{sec:Acknowledgements}Acknowledgements}
This manuscript has been authored by Fermi Research Alliance,
LLC under Contract No. DE-AC02-07CH11359 with the U.S. Department of Energy,
Office of Science, Office of High Energy Physics.
This work was supported by Brookhaven Science Associates, LLC under
Contract No. DESC0012704 with the U.S. Department of Energy.


\appendix

\newpage
\section{\label{secAPP:Radial}Parametrization of the radial part
of the map}

To determine the rotation number and parametrization of the radial
part of the map, we will use Danilov's theorem, as detailed in
\cite{zolkin2017rotation,nagaitsev2020betatron,zolkin2022mcmillan,
mitchell2021extracting}.
The steps to obtain $\nu_r$ are as follows:

\noindent
(1.) First, we introduce an auxiliary continuous system with
Hamiltonian identically equal to the radial invariant of the
map
\[
\K_r[\pr,r;t] =
    \pr^2 - a\,\pr\,r + r^2 \,\pmt\,\, \pr^2\,r^2 + \frac{\pt^2}{r^2}
\]
with the associated Hamilton's equations of motion:
\[
\begin{array}{l}
\ds \dot{r} = \frac{\pd \K}{\pd \pr} =
    2\,\pr\,(1\,\pmt\,\,r^2) - a\,r,        \\[0.35cm]
\ds -\dot{\pr} = \frac{\pd \K}{\pd r  } =
    2\,r\,(1\,\pmt\,\,\pr^2) - a\,\pr - \frac{2\,\pt^2}{r^3}.
\end{array}
\]

\noindent
(2.) Using the first Hamilton's equation, along with the
expression of momentum obtained from the invariant~(\ref{math:pr}),
we obtain:
\[
\dd t = \frac{\dd r}{2\,\pr\,(1\,\pmt\,\,r^2) - a\,r}
      = \frac{\pmv\,r\,\dd r}{2\,\sqrt{\mathcal{G}_6(r)}}
      = \frac{\pmv\,\dd \z}{4\,\sqrt{\mathcal{G}_3(\z)}}.
\]

\noindent
(3.) This allows us to write an expression for the period of
radial oscillations in the continuous system
\[
\T_r = \oint \, \dd t =
    \int_{\z_-}^{\z_+}\frac{\dd \z}{2\,\sqrt{\mathcal{G}_3(\z)}} =
    \frac{K[\kappa]}{\sqrt{\z_3-\z_1}}
\]
where the elliptic modulus $\kappa$ and complementary modulus
$\kappa' = \sqrt{1-\kappa^2}$ are given in terms of the roots
of $\mathcal{G}_3(\z)$
\[
\kappa = \kappa_\pmt
    \qquad\qquad\qquad
\kappa'= \kappa_\mpt
\]
with
\[
    \kappa_+ = \sqrt{\frac{\z_3-\z_2}{\z_3-\z_1}}
    \qquad\mathrm{and}\qquad
    \kappa_-= \sqrt{\frac{\z_2-\z_1}{\z_3-\z_1}}.
\]

\noindent
(4.) To relate the Hamiltonian to the radial part of the mapping,
we need to find the equivalent continuous-time interval for one
discrete step of the map.
This interval can be expressed as a one-step time integral, which
is given by
\begin{equation}
\label{math:T'}
\ds \T' = \frac{1}{2}\,\int_{r}^{r'}
    \frac{\pmv\,r\,\dd r}{\sqrt{\mathcal{G}_6(r)}}
        = \frac{1}{4}\,\int_{\z}^{\z'}
    \frac{\pmv\,\dd \z}{\sqrt{\mathcal{G}_3(\z)}},
\end{equation}
where the expression for $\z'$
\[
    \z' = \pr^2 \,\mpt\, \frac{\z_1\z_2\z_3}{\z},
\]
is obtained from the mapping equation and (\ref{math:ParamZ}).
It is worth noting that the integral~(\ref{math:T'}) is
independent of the initial value of $\z$, as long as it belongs
to the same level set of the invariant $\K_r = \const$.
A convenient choice of the initial point is one of the two stop
points, $r_\pm = \sqrt{\z_\pm}$.
Using the relations
\[
\pr\left[ r = r_\mp \right] =
    \frac{f(r)}{2}  =
    \frac{a}{2}\,\frac{r}{1\,\pmt\,r^2}
\]
and
\[
\pr^2\left[ \z = \z_\mp \right] =
    \frac{\z}{(1\,\pmt\,\z)^2}\,
    \prod\limits_{i=1}^3 (1 \,\pmt\,\z_i),
\]
we have
\[
\z_\mp \,\,\rightarrow\,\, \z_\mp':\qquad
\z_\mp' =
\frac{  \z_\mp(1 \,\pmt\, \z_\pm) \pm
        (\mpt\,\z_{2 \mpt 1})(\z_+-\z_-)}
{1 \,\pmt\, \z_\mp}
\]
and then evaluate the integral (\ref{math:T'}) as follows:
\[
\T' =
\left\{\begin{array}{ll}
    \T_\mu,        & a \geq 0,          \\[0.25cm]
    \T_r - \T_\mu, & a <    0,
\end{array}\right.
\]
where
\[
\T_\mu = \frac{1}{2\,\sqrt{\z_3-\z_1}}\,
    \eF\left[
        \arcsin \sqrt{\frac{\z_3-\z_1}{1\,\pmt\,\z_{2 \pmt 1}}},
        \kappa
    \right].
\]

\noindent
(5.) The previous two steps yield the expression for the radial
rotation number of the map
\[
\nu_r = \frac{\T'}{\T_r} =
\left\{\begin{array}{ll}
    \mu_r,     & a \geq 0,          \\[0.25cm]
    1 - \mu_r, & a <    0,
\end{array}\right.
\]
where
\[
\mu_r = \frac{1}{2\,\eK\left[\kappa\right]}\,
    \eF\left[
        \arcsin \sqrt{\frac{\z_3-\z_1}{1\,\pmt\,\z_{2 \pmt 1}}},
        \kappa
    \right].
\]
It should be noted that $\nu_r$ differs from the radial frequency
of the Hamiltonian system, which is given by
\[
\omega_r = \frac{2\,\pi}{\T_r} = \frac{2\,\pi}{\T'}\,\nu_r =
    \frac{2\,\pi\,\sqrt{\z_3-\z_1}}{\eK[\kappa]}.
\]

\noindent
(6.) Finally, by taking the integral~(\ref{math:T'}) from
$\z_0 = \z(0)$ to $\z = \z(t)$ and solving for $\z$, we get two
alternative forms of the solution:
\[
\begin{array}{rrr}
\z(t) =&\!  \z_\pmt \,\,\mpt\,\, (\z_+-\z_-)        &\!\!
    \sn^2\left[ 2\sqrt{\z_3-\z_1}\,t \,\mpt\, \phi_1,\kappa \right]      \\[0.25cm]
      =&\! \z_\mpt \,\,\pmt\,\, (\z_+-\z_-)         &\!\!
    \cn^2\left[ 2\sqrt{\z_3-\z_1}\,t \,\mpt\, \phi_1,\kappa \right]      \\[0.25cm]
      =&\! \z_{2 \mpt 1} \,\pmt\, (\z_3-\z_1)       &\!\!
    \dn^2\left[ 2\sqrt{\z_3-\z_1}\,t \,\mpt\, \phi_1,\kappa \right]
\end{array}
\]
or
\[
\begin{array}{rrr}
\z(t) =&\! \z_\pmt \,\mpt\quad\,\,\,\,\, (\z_+-\z_-)             &\!\!
\mathrm{cd}^2\left[ 2\sqrt{\z_3-\z_1}\,t \,\mpt\, \phi_2,\kappa \right]   \\[0.25cm]
      =&\! \z_\mpt \,\,\pmt\,\,\, \kappa'^2(\z_+-\z_-)    &\!\!
\mathrm{sd}^2\left[ 2\sqrt{\z_3-\z_1}\,t \,\mpt\, \phi_2,\kappa \right]   \\[0.25cm]
      =&\! \z_{2 \mpt 1} + (\z_\mpt-\z_{2 \mpt 1})  &\!\!
\mathrm{nd}^2\left[ 2\sqrt{\z_3-\z_1}\,t \,\mpt\, \phi_2,\kappa \right]
\end{array}
\]
where
\[
\begin{array}{l}
\ds \phi_1 = \pmv\,\eF\left[
    \arcsin \sqrt{\frac{\z_{2 \pmt 1}-\z_0}{\z_{2 \pmt 1}-\z_2}},
\kappa \right],                         \\[0.55cm]
\ds \phi_2 = \mpv\,\eF\left[
    \arcsin \frac{1}{\kappa}
    \sqrt{\frac{\z_2-\z_0}{\z_{2 \pmt 1}-\z_0}},
\kappa \right].
\end{array}
\]

The map's parametrization is given by:
\[
\begin{array}{ll}
\{r\}_n   &\!\!\!= \,r\,(n\,\T'),       \\[0.55cm]
\{\pr\}_n &\!\!\!= \pr  (n\,\T'),
\end{array}
\qquad\qquad
\begin{array}{ll}
    r  (t) &\!\!\!= \sqrt{\z(t)},           \\[0.35cm]
    \pr(t) &\!\!\!= \ds\frac{1}{2}\,\frac{\dot{r}(t)+a\,r(t)}
         {1\,\pmt\,r^2(t)},
\end{array}
\]
where
\[
\begin{array}{ll}
\ds \dot{r}(t) &\ds = \mpt\,2\,\kappa^2\,
        \frac{(\z_3-\z_1)^{3/2}}{r(t)}\,
        \sn_1\,\cn_1\,\dn_1                             \\[0.45cm]
&=\ds\mpt\,\frac{2}{\kappa}\,
        \frac{(\z_+-\z_-)^{3/2}}{r(t)}\,
        \mathrm{sd}_2\,\mathrm{cd}_2\,\mathrm{nd}_2,    \\[0.45cm]
\ds \dot{r}(t) &\ds = \pmt\,2\,(\kappa\,\kappa')^2\,
        \frac{(\z_3-\z_1)^{3/2}}{r(t)}\,
        \sn_1\,\cn_1\,\dn_1                             \\[0.45cm]
&=\ds\pmt\,\frac{2\,\kappa'^2}{\kappa}\,
        \frac{(\z_+-\z_-)^{3/2}}{r(t)}\,
        \mathrm{sd}_2\,\mathrm{cd}_2\,\mathrm{nd}_2,
\end{array}
\]
for the first or second form respectively.
In both cases, all elliptic functions have the same argument and
modulus as in the corresponding $\z(t)$.

\noindent
(7.) The action for the radial degree of freedom is given by the
same integral for both the mapping and the Hamiltonian:
\[
\begin{array}{ll}
\ds J_r &\ds = \frac{1}{2\,\pi}\,\oint \pr\,\dd r       \\[0.65cm]
        &\ds = \frac{1}{\pi}\,\int_{r_-}^{r_+}
            \frac{\sqrt{\mathcal{G}_6(r)}}{r\,(1\,\pmt\,r^2)}
        \,\dd r =
        \frac{1}{2\,\pi}\,\int_{\z_-}^{\z_+}
            \frac{\sqrt{\mathcal{G}_3(\z)}}{\z\,(1\,\pmt\,\z)}
        \,\dd \z                                        \\[0.65cm]
\ds     &\ds =
        \frac{1}{2\,\pi}\,\int_{\z_-}^{\z_+}\,\left[
        \frac{\z_1+\z_2+\z_3\,\pmt\,1}{\sqrt{\mathcal{G}_3(\z)}} -
        \frac{\z}{\sqrt{\mathcal{G}_3(\z)}}\,\pmt\right.\\[0.65cm]
\ds     &\ds \qquad\left.\pmt\,
        \frac{\z_1\,\z_2\,\z_3}{\z\,\sqrt{\mathcal{G}_3(\z)}} \,\mpt\,
        \frac{(1\,\pmt\,\z_1)(1\,\pmt\,\z_2)(1\,\pmt\,\z_3)}{(1\,\pmt\,\z)\,\sqrt{\mathcal{G}_3(\z)}}
        \right]\,\dd \z                                 \\[0.65cm]
\ds     &\ds  = \sqrt{\z_3-\z_1}\,\frac{\kappa'^2}{\pi}\,\left\{
        \pmt\,(1\,\pmt\,\z_{2\pmt1})\,\Pi\left[
            \kappa^2\,\frac{1\,\pmt\,\z_{2\mpt1}}{1\,\pmt\,\z_2},\kappa
        \right] - \right.                               \\[0.65cm]
\ds     &\ds \qquad\left.-\,
        \z_{2\pmt1}\,\Pi\left[
            \kappa^2\,\frac{\z_{2\mpt1}}{\z_2},\kappa
        \right] \,\mpt\,
        \Pi\left[ \kappa^2,\kappa \right]
        \right\}.
\end{array}
\]

\section{\label{secAPP:Angular}Parametrization of the angular part
of the map}

\noindent
In this section, we will discuss the steps required to obtain the
rotation number and parametrization for the angular variable
$\theta$.

\noindent
(1.) In order to apply Danilov theorem for the angular part of
the map we introduce a Hamiltonian
\[
    \h[\pr,r,\pt,\theta;t] = \K_r[\pr,r,\pt,\theta;t] + h(\pt)
\]
where we formally extend $\K_r$ to four dimensions and $h$ is an
unknown yet function that depends only on the angular momentum.
This Hamiltonian is separable in polar coordinates and has exactly
the same radial dynamics as $\K_r$.
The corresponding equations of motion for the angular variables
are:
\begin{equation}
\label{math:AngDifEq}
    \dot{\theta} = \frac{\pd \h}{\pd \pt} =
        \frac{2\,\pt}{r^2} + \varkappa
    \qquad\mathrm{and}\qquad
    \dot{\pt}    = -\frac{\pd \h}{\pd \theta} = 0,
\end{equation}
where $\varkappa = \pd h(\pt)/\pd t$ is a constant.
The Hamiltonian $\h$ preserves both $\K_r$ and $\pt$, and with
an appropriate choice of $h$, it should match the mapping equation
as:
\[
    \{\theta\}_n = \theta (n\,\T').
\]
Here, $\theta(t)$ is given by the integration
of~(\ref{math:AngDifEq})
\[
    \theta(t) = \theta_0 + \Theta(t) + \varkappa\,t
\]
 where $\Theta(t)$ is the solution of~(\ref{math:AngDifEq}) for
$h = 0$, and such that $\Theta(0)=0$.

\noindent
(2.) First, we solve for $\Theta(t)$ by setting $\varkappa=0$
in~(\ref{math:AngDifEq})
\begin{equation}
\label{math:dTheta}        
    \dd\Theta =
    \frac{2\,\pt}{r^2}\,\frac{\dd\,r}{2\,\pr\,(1\,\pmt\,\,r^2) - a\,r} =
    \frac{\pt}{2}\,\frac{\pmv\,\dd\z}{\z\,\sqrt{\mathcal{G}_3(\z)}}.
\end{equation}
We notice that derivative $\dot{\Theta}$ never vanishes unless
$\pt=0$, which implies that $\Theta(t)$ is a monotonically
increasing function of time for $\pt > 0$ (or decreasing for
$\pt < 0$) without turn or stop points.
Integrating~(\ref{math:dTheta}) from $r(0)=\{r\}_0$ to $r(t)$
we obtain:
\[
\begin{array}{l}
\ds\Theta(t) = \delta\Theta(t) - \delta\Theta(0),\\[0.3cm]
\ds\delta\Theta(t) =
    \pt\,\frac{
    \Pi\left[
        1-\frac{\z_\mpt}{\z_\pmt},
        \am\left[
            2\,\sqrt{\z_3-\z_1}\,t \,\mpt\, \phi_0,
            \kappa
        \right],
         \kappa
    \right]
    }
    {\z_\pmt\,\sqrt{\z_3-\z_1}}.
\end{array}
\]

\noindent
(3.) The function $\Theta(t)$ is arithmetic quasiperiodic, which
means that it satisfies the equation
\[
    \forall\,t:\qquad
    \Theta(t+\T_r) = \Theta(t) + \Delta_\Theta.
\]
In other words, it can be expressed as the sum of a periodic
function and a linear function, given by
\[
\Theta(t) = \Theta_\mathrm{per}(t) +
            \frac{\Delta_\Theta}{\T_r}\,t,
\qquad\qquad
\Theta_\mathrm{per}(t+\T_r) = \Theta_\mathrm{per}(t),
\]
where $\Theta_\mathrm{per}(t)$ is the periodic component of the
function and constant $\Delta_\Theta$ is a phase advance over one
radial oscillation
\[
\begin{array}{ll}
\ds\Delta_\Theta    &\ds = \oint \frac{2\,\pt}{r^2}\,\dd\,t     \\[0.35cm]
                    &\ds =
\pt\,\int_{\z_-}^{\z_+}\frac{\dd\z}{\z\,\sqrt{\mathcal{G}_3(\z)}} =
\frac{2\,\pt}{\z_\pmt\,\sqrt{\z_3-\z_1}}\,
    \Pi\left[
        1-\frac{\z_\mpt}{\z_\pmt},
        \kappa
    \right].
\end{array}
\]

\noindent
(4.) To relate the Hamiltonian $\h$ and angular map, we need to
determine the value of $\varkappa$.
We can do this by using the fact that the mapping equation relates
$\theta(\mathrm{T}')$ and $\theta(0)$, which gives us
\[
\Delta_\theta' \equiv
\theta(\mathrm{T}') - \theta(0) = \{\theta\}_0' - \{\theta\}_0
                       = \{\theta\}_1  - \{\theta\}_0.
\]
We can then express $\varkappa$ as
\[
    \varkappa = \frac{\Delta_\theta' - \Delta_\Theta'}{\mathrm{T}'}
\]
where $\Delta_\Theta'$ is the angular advance of $\Theta(t)$
over one step of the map, $\T'$:
\[
\Delta_\Theta' = \int_0^{\mathrm{T}'}\frac{2\,\pt}{r^2}\,\dd t
              = \int_{\z_0}^{\z_0'}\frac{\pt}{2}\,\frac{\pmv\,\dd\z}
                {\z\,\sqrt{\mathcal{G}_3(\z)}}.
\]

\noindent
(5.) The value of $\varkappa$ is independent of the initial
conditions and is determined solely by the values of $\pt$ and $\K$.
Both $\Delta_\theta'$ and $\Delta_\Theta'$, should be evaluated
from the same initial phase.
By choosing $r(0) = \{ r \}_0 = \sqrt{\z_\pmt}$, and using the
mapping equation, we obtain:
\[
\begin{array}{l}
\ds \Delta_\Theta' =
\left\{\begin{array}{ll}
    \Delta_\mu,                 & a \geq 0, \\[0.25cm]
    \Delta_\Theta - \Delta_\mu, & a <    0,
\end{array}\right.                                  \\[0.75cm]
\ds\Delta_\mu = \frac{\pt}{\z_\pmt\,\sqrt{\z_3 - \z_1}}\,
\Pi\left[
    \arcsin \sqrt{\frac{\z_3-\z_1}{1 \,\pmt\, \z_\pmt}},
    1-\frac{\z_\mpt}{\z_\pmt},
     \kappa
\right],                                            \\[0.75cm]
\ds \Delta_\theta'\,= 
\arctan\left(
    \frac{2\pt}{a}\,\frac{1\,\pmt\,\z_\pmt}{\z_\pmt}
\right) +
\pi\,\mathrm{sgn}[\pt]\,\mathrm{H}[-a].
\end{array}
\]

\noindent
(6.) We can now determine the action-angle variables.
For the angular degree of freedom, the action is simply the
absolute value of the angular momentum:
\[
    J_\theta = \frac{1}{2\,\pi}\,\oint \pt\,\dd \theta = |\pt|.
\]
The angular rotation number is defined as
\[
\nu_\theta \equiv \frac{\overline{\Delta_\theta'}}{2\,\pi}
                = \frac{\T'}{\T_r}\,
                  \frac{\overline{\Delta_\theta}}{2\,\pi}
\]
where $\overline{\Delta_\theta'}$ and $\overline{\Delta_\theta}$
are the averaged advances of the angular variable $\theta$ over
times $\T'$ and $\T_r$.
Since over one radial period the oscillatory part of $\theta(t)$
is averaged out, we have
\[
    \overline{\Delta_\theta} = 
    \Delta_\theta \equiv
    \theta(\T_r) - \theta(0)
\]
and thus
\[
\nu_\theta =
    \nu_r\,\frac{\Delta_\Theta + \varkappa\,\T_r}{2\,\pi} =
    \nu_r\,\frac{\Delta_\Theta}{2\,\pi} + \frac{\varkappa\,\T'}{2\pi} =
    \nu_r\,\frac{\Delta_\Theta}{2\,\pi} +
    \frac{\Delta_\theta'-\Delta_\Theta'}{2\,\pi}.
\]
The angular frequency for the Hamiltonian $\h$ is given by
\[
    \omega_\theta = \frac{\pd\h}{\pd J_\theta} =
    \frac{2\,\pi}{\T'}\,\nu_\theta =
    \frac{\Delta_\Theta}{\T_r} + \varkappa.
\]

\newpage
\section{Special functions and integrals.}
\label{secAPP:Integrals}

\subsection{Normal elliptic integrals}

The set of three fundamental integrals
\[
\begin{array}{ll}
\eF[\phi,k]               &
\ds =   \int_0^\phi \frac{1}{\sqrt{1-k^2\sin^2\theta}}\,\dd\theta \\[0.5cm]
&\ds=   \int_0^x    \frac{1}{\sqrt{(1-t^2)(1-k^2\,t^2)}}\,\dd t,  \\[0.5cm]
\eE[\phi,k]               &
\ds =   \int_0^\phi \sqrt{1-k^2\sin^2\theta}\,\,\dd\theta         \\[0.5cm]
&\ds=   \int_0^x    \sqrt{\frac{1-t^2}{1-k^2\,t^2}}\,\dd t,       \\[0.5cm]
\Pi[\phi,\alpha^2,k]    &
\ds =   \int_0^\phi \frac{1}{(1-\alpha^2\sin^2\theta)
                        \sqrt{1-k^2\sin^2\theta}}\,\dd\theta      \\[0.5cm]
&\ds=   \int_0^x    \frac{1}{(1-\alpha^2 t^2)
                             \sqrt{(1-t^2)(1-k^2\,t^2)}}\,\dd t,
\end{array}     
\]
are called {\it incomplete elliptic integrals of the first},
{\it the second} and {\it the third kind} respectively.
They are functions of two arguments: the amplitude $\phi$ and
elliptic modulus, or simply the modulus, $k$.
In addition, the third integral depends on argument $\alpha$
called the {\it characteristic} with $-\infty < \alpha^2 < \infty$.
The first and the second forms of integrals are related through the
change of variables
\[
x = \sin\phi
\qquad\qquad\mathrm{and}\qquad\qquad
t = \sin\theta.
\]
They are known as {\it Legendre’s} and {\it Jacobi’s canonical
forms}.
When the amplitude $\phi = \pi/2$ ($x=1$), the integrals are said
to be {\it complete} and denoted as
\[
\begin{array}{l}
\ds \eK[k] = \eF[\pi/2,k],      \\[0.3cm]
\ds \eE[k] = \eE[\pi/2,k],      \\[0.3cm]
\ds \Pi[\alpha^2,k] = \Pi[\pi/2,\alpha^2,k].
\end{array}
\]

\subsection{Jacobi's amplitude function}

The {\it Jacobi's amplitude} or simply {\it amplitude} function
can be defined as the inverse of the incomplete elliptic integral
of the first kind
\[
    \am(t,k) = \phi
    \qquad\mathrm{where}\qquad\qquad
    t = \eF[\phi,k].
\]
$\am(t,k)$ is a monotonic infinitely differentiable function of
$t$ with special values
\[
    \am\,(0,k) = 0,
    \qquad\qquad
    \am\,(\eK[k],k) = \pi/2,
\]
and {\it arithmetic quasiperiodicity}
\[
    \forall\,t:\qquad
    \am(t+2\,\eK[k],k) = \am(t,k) + \pi.
\]

\subsection{Jacobi elliptic functions}

In general there are 12 Jacobi elliptic functions which are
related in the following way.
Let $p$, $q$ and $r$ be any three of the letters $s$, $c$, $d$ and
$n$.
Then, with the convention $\mathrm{pp}=\mathrm{qq}=\mathrm{rr}=1$,
{\it Glaisher's notation} holds
\[
\mathrm{pq}(t,k) = \frac{\mathrm{pr}(t,k)}{\mathrm{qr}(t,k)}
                 = \frac{1}{\mathrm{qp}(t,k)}.
\]
Three primary functions, {\it elliptic sine} $\sn$, {\it elliptic
cosine} $\cn$ and {\it delta amplitude} $\dn$, can be defined
using Jacobi's amplitude $\am$
\[
\begin{array}{l}
    \sn(t,k) = \sin\phi = \sin[\am(t,k)],   \\[0.4cm]
    \cn(t,k) = \cos\phi = \cos[\am(t,k)]
\end{array}
\]
and relation between squares of the functions
\[
\begin{array}{l}
    \cn^2(t,k) + \sn^2(t,k) = 1,    \\[0.4cm]
    \cn^2(t,k) + k'^2\,\sn^2(t,k) = \dn^2(t,k).
\end{array}
\]
All three functions a periodic with periods equal to $4\,\eK[k]$
for $\sn$ and $\cn$, and $2\,\eK[k]$ for the delta amplitude $\dn$.

\subsection{List of integrals involving elliptic functions}

Below, for the convenience of the readers, we provide a list of
integrals used in this article.
Introducing
\[
\kappa = \sqrt{\frac{\z_3-\z_2}{\z_3-\z_1}}
\qquad\text{and}\qquad
\kappa'= \sqrt{\frac{\z_2-\z_1}{\z_3-\z_1}}
\]
along with
\[
\begin{array}{l}
\ds\phi_1 = \arcsin\sqrt{\frac{\z_3-\z}{\z_3-\z_2}},                \\[0.6cm]
\ds\phi_2 = \arcsin\frac{1}{\kappa}\,\sqrt{\frac{\z-\z_2}{\z-\z_1}},\\[0.6cm]
\ds\phi_3 = \arcsin\frac{1}{\kappa}\,\sqrt{\frac{\z_2-\z}{\z_3-\z}},\\[0.6cm]
\ds\phi_4 = \arcsin\sqrt{\frac{\z-\z_1}{\z_2-\z_1}},
\end{array}
\]

\newpage

\noindent
if $\z_1 < \z_2 \leq \z \leq \z_3$:
\[\begin{array}{l}
\ds \int_\z^{\z_3}
\frac{\dd\z}{\sqrt{\mathcal{G}_3(\z)}} =
    2\,\frac{\eF[\phi_1,\kappa]}{\sqrt{\z_3-\z_1}},     \\[0.65cm]
\ds \int_{\z_2}^\z
\frac{\dd\z}{\sqrt{\mathcal{G}_3(\z)}} =
    2\,\frac{\eF[\phi_2,\kappa]}{\sqrt{\z_3-\z_1}},     \\[0.65cm]
\ds \int_\z^{\z_3}
\frac{\z\,\dd\z}{\sqrt{\mathcal{G}_3(\z)}} =
    2\,\frac{\z_1\,\eF[\phi_1,\kappa] +
        \frac{\z_3}{\z_2}\,(\z_3-\z_1)\,\eE[\phi_1,\kappa]}
        {\sqrt{\z_3-\z_1}}                              \\[0.65cm]
\ds \int_{\z_2}^\z
\frac{\z\,\dd\z}{\sqrt{\mathcal{G}_3(\z)}} =
    2\,\frac{\z_1\,\eF[\phi_2,\kappa] +
        (\z_2-\z_1)\,\Pi[\phi_2,\kappa^2,\kappa]}
        {\sqrt{\z_3-\z_1}},                             \\[0.65cm]
\ds \int_\z^{\z_3}
\frac{\dd\z}{(\z-p)\,\sqrt{\mathcal{G}_3(\z)}} =
    2\,\frac{\Pi\left[\phi_1,\frac{\z_3-\z_2}{\z_3-p},\kappa\right]}
    {(\z_3-p)\,\sqrt{\z_3-\z_1}},                       \\[0.65cm]
\ds \int_{\z_2}^\z
\frac{\dd\z}{(\z-p)\,\sqrt{\mathcal{G}_3(\z)}} =
    2\times                                             \\[0.65cm]
\ds \qquad\qquad
    \frac{\eF[\phi_2,\kappa] -
        \frac{\z_2-\z_1}{\z_2-p}\,\Pi\left[
            \phi_2,\kappa^2\,\frac{\z_1-p}{\z_2-p},\kappa
        \right]}{(\z_1-p)\,\sqrt{\z_3-\z_1}},
\end{array}\]
and if $\z_1 \leq \z \leq \z_2 < \z_3$:
\[\begin{array}{l}
\ds \int_\z^{\z_2}
\frac{\dd\z}{\sqrt{\mathcal{G}_3(\z)}} =
    2\,\frac{\eF[\phi_1,\kappa]}{\sqrt{\z_3-\z_1}},     \\[0.65cm]
\ds \int_{\z_1}^\z
\frac{\dd\z}{\sqrt{\mathcal{G}_3(\z)}} =
    2\,\frac{\eF[\phi_2,\kappa]}{\sqrt{\z_3-\z_1}},     \\[0.65cm]
\ds \int_\z^{\z_2}
\frac{\z\,\dd\z}{\sqrt{\mathcal{G}_3(\z)}} =
    2\,\frac{\z_3\,\eF[\phi_1,\kappa] -
        (\z_3-\z_2)\,\Pi[\phi_1,\kappa^2,\kappa]}
        {\sqrt{\z_3-\z_1}},                             \\[0.65cm]
\ds \int_{\z_1}^\z
\frac{\z\,\dd\z}{\sqrt{\mathcal{G}_3(\z)}} =
    2\,\frac{\z_3\,\eF[\phi_2,\kappa] -
        (\z_3-\z_1)\,\eE[\phi_2,\kappa]}
        {\sqrt{\z_3-\z_1}},                             \\[0.65cm]
\ds \int_\z^{\z_2}
\frac{\dd\z}{(p-\z)\,\sqrt{\mathcal{G}_3(\z)}} =
    2\times                                             \\[0.65cm]
\ds \qquad\qquad
    \frac{\eF[\phi_1,\kappa] -
        \frac{\z_3-\z_2}{p-\z_2}\,\Pi\left[
            \phi_1,\kappa^2\,\frac{p-\z_3}{p-\z_2},\kappa
        \right]}{(p-\z_3)\,\sqrt{\z_3-\z_1}},           \\[0.65cm]
\ds \int_{\z_1}^\z
\frac{\dd\z}{(p-\z)\,\sqrt{\mathcal{G}_3(\z)}} =
    2\,\frac{\Pi\left[\phi_2,\frac{\z_2-\z_1}{p-\z_1},\kappa\right]}
    {(p-\z_1)\,\sqrt{\z_3-\z_1}}.
\end{array}\]

\newpage

%

\end{document}